\documentclass[twocolumn]{aastex701}
\usepackage{graphicx}
\usepackage{amsmath}
\usepackage{booktabs}
\usepackage{comment}
\usepackage{makecell}
\usepackage{array}
\usepackage{soul}
\usepackage{todonotes}
\usepackage{float}
\usepackage[shortlabels]{enumitem}
\usepackage{algorithm}
\usepackage{algpseudocode}
\usepackage{rotating}

\received{}
\revised{}
\accepted{}

\submitjournal{one of the AAS Journals}

\shorttitle{A Data-Driven Iterative Algorithm for Spectral Line Selection}
\shortauthors{K. Pandey, et al.}

\newcommand{\PSUAA}{Department of Astronomy and Astrophysics, 525 Davey Laboratory, 251 Pollock Road, Penn State, University Park, PA, 16802, USA}
\newcommand{\PSUCEHW}{Center for Exoplanets and Habitable Worlds, 525 Davey Laboratory, 251 Pollock Road, Penn State, University Park, PA, 16802, USA}
\newcommand{\PSUARC}{Astrobiology Research Center, 525 Davey Laboratory, 251 Pollock Road, Penn State, University Park, PA, 16802, USA}
\newcommand{\PSETI}{Penn State Extraterrestrial Intelligence Center, 525 Davey Laboratory, 251 Pollock Road, Penn State, University Park, PA, 16802, USA}
\newcommand{\UA}{Steward Observatory, University of Arizona, 933 N.\ Cherry Ave, Tucson, AZ 85721, USA}

\newcommand{\Penn}{Department of Physics and Astronomy, University of Pennsylvania, 209 S 33rd St, Philadelphia, PA 19104, USA}

\newcommand{\NOAO}{U.S. National Science Foundation National Optical-Infrared Astronomy Research Laboratory, 950 N.\ Cherry Ave., Tucson, AZ 85719, USA}
\newcommand{\UW}{Department of Statistics, University of Wisconsin, 1205 University Avenue, Madison, WI 53706, USA}
\newcommand{\Macquarie}{School of Mathematical and Physical Sciences, Macquarie University, Balaclava Road, North Ryde, NSW 2109, Australia}

\newcommand{\UCI}{Department of Physics \& Astronomy, The University of California, Irvine, Irvine, CA 92697, USA}
\newcommand{\Carleton}{Carleton College, One North College St., Northfield, MN 55057, USA}

\newcommand{\PSUICS}{Institute for Computational and Data Sciences, Penn State, University Park, PA, 16802, USA}
\newcommand{\PSUCASt}{Center for Astrostatistics, 525 Davey Laboratory, 251 Pollock Road, Penn State, University Park, PA, 16802, USA}

\newcommand{\FlatironCCA}{Center for Computational Astrophysics, Flatiron Institute, 162 Fifth Avenue, New York, NY 10010, USA}

\newcommand{\TIFR}{Department of Astronomy and Astrophysics, Tata Institute of Fundamental Research, Homi Bhabha Road, Colaba, Mumbai 400005, India}

\newcommand{\Schmidt}{Astrophysics \& Space Institute, Schmidt Sciences, New York, NY 10011, USA}
\newcommand{\UChicago}{Department of Astronomy \& Astrophysics, University of Chicago, 5801 South Ellis Avenue, Chicago, IL 60637, USA}

\begin{document}

\title[\LARGE Improving the Precision of Line-by-Line Radial Velocities: A Data-Driven Iterative Algorithm for Spectral Line Selection]{Improving the Precision of Line-by-Line Radial Velocities: A Data-Driven Iterative Algorithm for Spectral Line Selection}

\author[0000-0001-9943-5538]{Kanishk Pandey}
\affiliation{\PSUAA}
\affiliation{\PSUCEHW}
\affiliation{\PSUICS}
\email{kfp5432@psu.edu}

\author[0000-0001-6545-639X]{Eric Ford}
\affiliation{\PSUAA}
\affiliation{\PSUCEHW}
\affiliation{\PSUICS}
\affiliation{\PSUCASt}
\email{ebf11@psu.edu}

\author[0009-0009-4165-9606]{Joseph M. Salzer} 
\affiliation{\UW}
\email{jsalzer@wisc.edu}

\author[0000-0002-9656-2272]{Jessi Cisewski-Kehe} 
\affiliation{\UW}
\email{jjkehe@wisc.edu}

\author[0000-0002-3852-3590]{Lily L. Zhao}
\altaffiliation{Sagan Fellow}
\affiliation{\FlatironCCA}
\affiliation{\UChicago}
\email{lilylingzhao@uchicago.edu}

\author[0000-0002-5463-9980]{Arvind F.\ Gupta}
\affiliation{\NOAO}
\email{arvind.gupta@noirlab.edu}

\author[0000-0002-4677-8796]{Michael L. Palumbo III}
\affiliation{\FlatironCCA}
\email{mpalumbo@flatironinstitute.org}

\author[0000-0002-4788-8858]{Ryan C. Terrien}
\affiliation{\Carleton}
\email{rterrien@carleton.edu}

\author[0000-0003-0199-9699]{Evan Fitzmaurice}
\affiliation{\PSUAA}
\affiliation{\PSUCEHW}
\affiliation{\PSUICS}
\email{exf5296@psu.edu}

\author[0009-0000-2099-6130]{Brady Barth}
\affiliation{\PSUAA}
\email{bjb6796@psu.edu}

\author[0000-0001-5137-1176]{Jason T. Wright}
\affiliation{\PSUAA}
\affiliation{\PSUCEHW}
\affiliation{\PSETI}
\email{astrowright@gmail.com}

\author[0000-0002-6096-1749]{Cullen H.\ Blake}
\affiliation{\Penn}
\email{chblake@sas.upenn.edu}

\author[0000-0001-9596-7983]{Suvrath Mahadevan}
\altaffiliation{Principal Investigator}
\affiliation{\PSUAA}
\affiliation{\PSUCEHW}
\affiliation{\PSUARC}  
\email{suvrath@astro.psu.edu}

\author[0000-0001-8127-5775]{Arpita Roy}
\affiliation{\Schmidt}
%\affiliation{\STScI}
%\affiliation{\JHU}
\email{arpita308@gmail.com}

\author[0000-0001-8720-5612]{Joe P.\ Ninan}
\affiliation{\TIFR}
%\affiliation{\PSUAA}
%\affiliation{\PSUCEHW}
\email{indiajoe@gmail.com}

\author[0000-0001-7409-5688]{Gudmundur Stefansson}
\affiliation{\Schmidt}
\affiliation{Anton Pannekoek Institute for Astronomy, University of Amsterdam, Science Park 904, 1098 XH Amsterdam, The Netherlands}
\email{g.k.stefansson@uva.nl}

\author[0000-0003-0149-9678]{Paul Robertson}
\altaffiliation{Instrument Team Project Scientist}
\affiliation{\UCI}
\email{paul.robertson@uci.edu}

\author[0000-0002-4046-987X]{Christian Schwab}
\affiliation{\Macquarie}
\email{mail.chris.schwab@gmail.com}

\author[0000-0001-9626-0613]{Daniel M.\ Krolikowski}
\affiliation{\UA}
\email{krolikowski@arizona.edu}

\author[0000-0003-1324-0495]{Leonardo A.\ Paredes}
\affiliation{\UA}
\email{lparedes@arizona.edu}

\correspondingauthor{Kanishk Pandey}
\email{kfp5432@psu.edu}

\begin{abstract}
Independent analysis of individual spectral lines, or line-by-line (LBL) analyses, can improve upon standard cross-correlation function (CCF) methods for measuring radial velocities (RVs) because they preserve critical information about individual line shape changes that can be caused by stellar activity. 
In this work, we measure LBL RVs of 3,830 spectral lines across 383 days of NEID solar observations. 
Our LBL approach achieves an RV RMS of $2.012~\mathrm{m\,s^{-1}}$, which is slightly lower than the $2.129~\mathrm{m\,s^{-1}}$ achieved by a CCF approach using a shared line list.  
Then, we describe and benchmark several methods for selecting line lists based on line properties such as depth and intrinsic RV scatter. 
We find that these subsets have a lower RV RMS compared to either the full line list or random subsets of equal size.
Motivated by these results, we present FLARES (Filtering Lines for Accurate Radial-velocity Exoplanet Search), an iterative line-selection algorithm.
FLARES selects candidate spectral lines with extreme values of multiple line metrics and properties such as depth, signal-to-noise ratio, and detector position, and preferentially rejects lines whose removal produces the largest decrease in the weighted RV scatter.
FLARES achieves an RV RMS of $1.122~\mathrm{m\,s^{-1}}$ using just 24 lines and performs better than the benchmark methods. 
We perform Monte Carlo simulations and show FLARES is robust and reproducible.
Comparisons to alternative line lists chosen to have properties similar to the best FLARES-selected lines demonstrate that FLARES is successfully identifying line properties that lead to effective line lists for future extreme-precision RV measurements.  
\end{abstract}


\section{Introduction} \label{Introduction}
Since the discovery of the first exoplanet \citep{mayor_queloz_1995}, thousands of exoplanets have been discovered through a variety of methods \citep{winn_fabrycky_2015}. 
The radial velocity (RV) technique alone accounts for about $20\%$ of the over 6,000 exoplanets that have been confirmed to date \citep{burt_2025, christiansen_2025}, and is the most prolific technique for constraining planetary masses. 
However, an Earth-mass planet orbiting in the Habitable Zone of a Sun-like star only results in an RV amplitude of $\approx$10 cm~s$^{-1}$, which requires extreme precision radial velocity (EPRV) measurements. 

Although there has been great progress in the precision of highly-stabilized EPRV spectrographs \citep{pepe_2004, cosentino_2012, pepe_2013, schwab_2016, jurgenson_2016, seifahrt_2018, gibson_2024}, detecting planets with RV amplitudes at the sub m~s$^{-1}$ level is difficult.
For example, the nominal resolving power of the NEID spectrograph \citep{schwab_2016} is R = $\lambda / \Delta \lambda \approx 120,000$, which means the effective 1-dimensional (1D) spectral line spread function (LSF) width spans about $\approx 2,500$ m~s$^{-1}$. 
Assuming a 5-pixel sampling of the LSF for NEID \citep{schwab_2016}, this equates to an effective velocity dispersion of 500 m~s$^{-1}$ per pixel. 
Therefore, a signal of 10 cm~s$^{-1}$ from an Earth-like planet around a Sun-like star would only shift the stellar spectrum by about 0.0002 pixels. 

An additional complication is that stellar variability can introduce quasi-periodic RV variations, often greater than $\sim$1 m~s$^{-1}$ \citep{dumusque_2011b, dumusque_2011a, cretignier_2020, luhn_2020, burt_2025}.
Signals from pressure-mode (p-mode) oscillations ($\sim 5$ min. timescales, $\sim$10$-$100 cm~s$^{-1}$ RV jitter; \citealp{kjeldsen_bedding_1995, burt_2025}), granulation ($\sim$ minute-to-hour timescales, $\sim$0.3$-$0.4 m s$^{-1}$ RV jitter; \citealp{dravins_1981}) and supergranulation ($\sim$ day timescales, $\sim$0.7$–$0.9 m~s$^{-1}$ RV jitter; \citealp{al_moulla_2023, lakeland_2024, OSullivan_2025}) can obscure the RV signal from exoplanets and either limit the effective detection sensitivity of EPRV surveys or lead to potential false-positive detections \citep{crass_2021}. 

\begin{comment}
An additional complication is that stellar variability can introduce quasi-periodic RV variations, often greater than $\sim$1 m~s$^{-1}$ \citep{dumusque_2011b, dumusque_2011a, cretignier_2020, luhn_2020, burt_2025}.
These stellar signals can obscure the RV signal from exoplanets either limiting the effective detection sensitivity of EPRV surveys or leading to potential false-positive detections \citep{crass_2021}. 
For example, pressure mode (p-mode) oscillations are standing acoustic waves at timescales of about five minutes for the Sun and can introduce RV jitter on the order of $\sim$10$-$100 cm~s$^{-1}$ \citep{kjeldsen_bedding_1995, burt_2025}. 
Granulation refers to the process whereby hot plasma in convective cells rises to the stellar surface with high velocities and sinks back into the intergranular lanes between these cells on minute-to-hour timescales \citep{dravins_1981}. 
These cells can reach velocities of $\sim$1 km~s$^{-1}$, though they only leave a residual RV jitter of $\sim$0.3$-$0.4 m s$^{-1}$ after they are averaged over the surface of the Sun  \citep{dumusque_2011a, al_moulla_2023}. 
Supergranulation occurs for much larger cells and operates on longer timescales of a few days, contributing a further $\sim$0.7$–$0.9 m~s$^{-1}$ of RV variability \citep{al_moulla_2023, lakeland_2024, OSullivan_2025}.
\end{comment}

Magnetic activity also plays a crucial role in stellar variability. 
Dark starspots and faculae (bright, active regions) can alter the surface flux distribution and break the symmetry between the blueshifted and redshifted sides of the stellar rotation profile \citep{saar_donahue_1997, cavallini_1985}. 
As these features rotate into and out of view, they induce RV variations at the stellar rotation period \citep{saar_donahue_1997, meunier_2010}. 
In addition, magnetic fields inhibit convective blueshift, causing spots and faculae to appear redshifted relative to the quiet photosphere and generate a net positive RV shift \citep{dravins_1982}.
Collectively, these effects can produce spurious stellar jitter at sub-m~s$^{-1}$ level in quiet, slowly rotating stars \citep{meunier_2010} to several tens of m~s$^{-1}$ in young or magnetically active stars \citep{boisse_2011, dumusque_2014}.
Finally, long-term magnetic activity cycles can change the filling factor of active regions and produce long-term RV shifts that correlate with activity indicators \citep{lovis_2011}. 

The Sun offers a unique opportunity to understand the effect of stellar activity signals on RV measurements \citep{mcmillan_1993, butler_1996, saar_2003, lagrange_2010, meunier_2010, lagrange_2011, haywood_2016, lanza_2016, wright_2020}.  First, since the true center-of-mass shifts due to planets are known for the Sun, it is possible to precisely characterize spurious RVs introduced by stellar variability without confusions from potential undetected planets \citep{wright_2020}. 
Second, since the Sun is a spatially resolved source, it is possible to empirically characterize how different sources of variability, such as spots and faculae, can influence measured RVs.
Finally, the Sun's brightness and daytime observability enable high-cadence observations and extremely high signal-to-noise ratio (SNR) spectra without conflicting with night-time observations. 
This allows for substantial averaging of short-timescale variability such as p-mode oscillations and granulation \citep{dumusque_2011a, meunier_2015}.

%different physical mechanisms such as spots and faculae influence measured RVs using software like the Spot Oscillation And Planet code (SOAP, \citealp{boisse_2012, dumusque_2014, akinsanmi_2018, cristo_2025}).

\begin{comment}
There are also practical advantages of solar observations.
For example, the Sun is observable during the day which means it is practical to take hundreds of exposures per clear day without conflicting with night-time observations. 
The individual day-time exposures can be combined to give high signal-to-noise ratio (SNR) daily-averaged solar spectra. 
Short-timescale oscillations such as p-modes and granulation can be mitigated in both solar and stellar observations through longer integrations or by averaging multiple consecutive exposures \citep{dumusque_2011a}. 
However, the exceptionally high cadence achievable for solar observations also enables substantial averaging down of granulation signals through daily averaging, which is generally impractical for other stars due to limited observing time \citep{meunier_2015}. 
\end{comment}

In addition, nearly contemporaneous observations from multiple EPRV spectrographs such as HARPS, HARPS-N, EXPRES, and NEID suggest that each pair of instruments largely agreed to within ~$15-30$ cm~s$^{-1}$ \citep{zhao_2023}. Each instrument generally follows similar long-term trends throughout the month, with ~$50-60$ cm~s$^{-1}$ scatter day to day, though some of this scatter can be explained due to the differential effect of the Sun rising and setting over different instruments \citep{zhao_2023}. The similarity of precise RVs from NEID and \mbox{HARPS-N} was also shown in \citet{dumusque_2025}, where they found that over a three year period, the scatter of the differences between the RVs of both instruments was 0.79 m~s$^{-1}$. 
During the low activity phase, the scatter was only 0.6 m~s$^{-1}$ which is expected from granulation and the different longitudes of the two instruments \citep{lakeland_2024}. 

However, there are also challenges associated with observing the Sun. 
Since the Sun is a resolved source, any differences in airmass, cloud coverage, and atmospheric distortions across the disk of the Sun can result in spurious RV signals \citep{collier_cameron_2019}. 
In addition, the Earth–Sun relative velocity is small, which means telluric absorption lines shift little relative to solar lines and make them hard to separate \citep{zhao_2023}. 

\begin{comment}
The Extreme Stellar Signals Project (ESSP) analyzed Sun-as-a-star observations from four high resolution spectrographs: HARPS, \mbox{HARPS-N}, EXPRES, and NEID, in order to gain insight on the RV precision and accuracy of current generation of EPRV spectrographs \citep{zhao_2023}. 
After binning exposures from each instrument within a day, each pair of instruments largely agreed to within ~$15-30$ cm~s$^{-1}$. 
In addition, each instrument generally followed similar long-term trends throughout the month, with ~$50-60$ cm~s$^{-1}$ scatter day to day, though some of this scatter can be explained due to the differential effect of the Sun rising and setting over different instruments. 
Indeed, differential extinction can cause spurious RV signals of 50 cm~s$^{-1}$ at an airmass of 2.2 \citep{collier_cameron_2019, zhao_2023}. 
\end{comment}

\begin{comment}
The similarity of precise RVs from NEID and \mbox{HARPS-N} was also shown in \citet{dumusque_2025}, where they found that over a three year period, the scatter of the differences between the RVs of both instruments was 0.79 m~s$^{-1}$. 
During the low activity phase, the scatter was only 0.6 m~s$^{-1}$ which is expected from granulation and the different longitudes of the two instruments. 
However, there were some periods showing larger RV differences.   
One possible explanation is that these differences arise due to the different weighting scheme for different orders applied in \citet{ford_2024} which could be less sensitive to activity. 
\end{comment}

Despite the high precision achievable with modern EPRV instruments, stellar variability and instrumental effects continue to limit the precision of RV measurements. 
This motivates the use of advanced techniques to extract Doppler shifts from stellar spectra. 
One widely used approach is the cross-correlation function (CCF) method, in which an observed stellar spectrum is cross-correlated with a line mask composed of weighted absorption features. 
The resulting CCF (as a function of the velocity shift of the line mask) represents an average line profile, with the RV estimated from the location of its minimum, typically obtained via a Gaussian fit \citep{baranne_1996, pepe_2002}. 
Later improvements introduced depth/SNR-based weighted masks, which improved sensitivity and robustness \citep{pepe_2002, lafarga_2021}. 
An alternative approach is based on template matching, where the RV is estimated by how much each individual observation needs to be shifted to best match a template (e.g., a high S/N spectrum constructed by co-adding observations; \citealp{butler_1996, anglada_escuda_tuomi_2012, zechmeister_2018, gilbertson_2024}).

In parallel to the CCF-based and template-matching approaches, there has also been a push to derive line-by-line (LBL) RVs using a template-based approach or Gaussian fits to each line.
Although each individual LBL RV has relatively low precision, a weighted average of many LBL RV measurements can provide a single precise RV. 
This has the advantage of making it practical to remove, or even correct for, the lines that are particularly susceptible to stellar activity signals or blends. 
In addition, traditional CCF and template matching approaches aggregate all line information into one RV which ``averages out" line shape information. 
In contrast, a LBL approach can preserve line shape information which can be useful for determining which lines are most susceptible to stellar activity and/or removing some of the stellar activity signals.

Many LBL approaches have had success in mitigating the effects of stellar activity. 
\citet{dumusque_2018} measured LBL RVs from HARPS spectra and demonstrated close agreement with CCF-based RVs. 
They further showed that selecting activity-insensitive lines can reduce the RV RMS of $\alpha$ Cen B by a factor of 1.6, though with increased white noise from using fewer lines.
\citet{cretignier_2020} found the effects of activity-driven convective blueshift in individual LBL RVs from $\alpha$ Cen B spectra, with shallower lines exhibiting greater sensitivity. 
They reduced the signal to $\sim$0.8 m s$^{-1}$ using depth-dependent line selection and decorrelation techniques. \citet{cretignier_2021} reduced the RV RMS of Tau Ceti from 1.18 to 1.02 m s$^{-1}$ using a suite of post-processing techniques. Their LBL approach decorrelated RVs using the depth and fluxes on the left and right sides of the wings to suppress activity-induced RV signals.
\citet{almoulla_2022} studied how stellar activity signals depend on spectral line formation temperature by constructing RV time series from different temperature bins. They found that the amplitude, phase, and dominant timescales of activity signals vary with formation temperature, with hotter regions more strongly tracing chromospheric activity and convective suppression than cooler lines.
Using simulations from the GRanulation And Spectrum Simulator (GRASS), \citet{palumbo_2024} found that different spectral lines exhibit a range of approximately 0.3 m s$^{-1}$ in granulation-induced RV scatter. Their results suggest that selecting intrinsically stable lines may help mitigate granulation-driven RV variability.
\citet{salzer_2025} used a fixed effects linear model to estimate RV signals by using line shape changes across time to disentangle stellar activity signals from true Doppler shifts. 
Including shape changes in the model reduced the RV RMS from $\sim$1.7 m s$^{-1}$ to $\sim$0.4 m s$^{-1}$.

Many LBL studies demonstrate that different lines respond differently to stellar activity and that exploiting these differences can improve RV precision. 
Motivated by these results, we introduce a new algorithm that iteratively removes spectral lines from a full line list in order to identify subsets of lines that are less sensitive to stellar variability, instrumental systematics, and telluric contamination and demonstrate this algorithm on NEID solar data.
Section \ref{Data} describes the NEID solar observations used in this work, along with details about the line list, line fitting and LBL RV estimation, and the procedure to remove lines showing spurious RVs. 

Section \ref{Benchmark Line Selection Algorithms} describes several benchmark line selection algorithms that are a useful comparison with later methods described in this work. 
Section \ref{Filtering Lines for Accurate RV Exoplanet Surveys (FLARES)} describes an iterative filtering algorithm that uses multiple metrics to select candidate lines for removal from a full line list. 
Section \ref{Results} describes the results of this approach including the metrics most influential to the algorithm, the properties of the best lines, ways to quantify the robustness and reproducibility of the algorithm using Monte Carlo simulations, and developing ensembles of comparison line lists with similar physical properties as the best chosen lines. 
Finally, Section \ref{Conclusions} summarizes our conclusions and discusses possible directions for future work. 
We also release a dataset of RVs (and corresponding uncertainties) for the curated line list (based on Section \ref{Line Vetting}) across 383 days of NEID solar observations, the metric values for the lines, and a LBL ranking list based on algorithms introduced in this work on Zenodo \citep{zenodo_pandey_2026}. 

\section{Data} \label{Data}

\subsection{NEID Solar Data \label{NEID Solar Data}}

We use solar observations from NEID, which is a high-precision, ultra-stabilized, fiber-fed spectrograph designed to conduct EPRV measurements  \citep{schwab_2016}. 
NEID is located at Kitt Peak National Observatory in Arizona and receives starlight from the WIYN \footnote{The WIYN Observatory is a joint facility of the NSF’s National Optical-Infrared Astronomy Research Laboratory, Indiana University, the University of Wisconsin–Madison, Pennsylvania State University and Princeton University.} 3.5-m Telescope. 
NEID is designed to have a single-measurement instrumental precision that exceeds 27 cm~s$^{-1}$ \citep{halverson_2016}. 

An auxiliary NEID solar telescope enables daytime observations of the Sun with 55 s exposures and 28 s readout time from 16:30 to 22:30 UT \citep{lin_2022}. 
This yields $\sim$200 observations per day. 
These high-cadence data can be binned to achieve higher SNR as well as mitigate the effects of short-timescale stellar variability such as p-mode oscillations and granulation \citep{dumusque_2011a, collier_cameron_2019}.
These solar data are collected at a resolution of $\sim$120,000 over a broad wavelength range (380–930 nm) and use a Laser Frequency Comb (LFC), hollow-cathode lamp (Thorium-Argon, with Uranium-Neon beginning in August 2023)\footnote{\url{https://neid.ipac.caltech.edu/docs/NEID-DRP/algorithms.html\#daily-wavelength-calibration}}, and Fabry-Pérot etalon for wavelength calibration. 
Early NEID solar observations demonstrate an RV RMS of 0.66 m~s$^{-1}$ under good sky conditions \citep{lin_2022} and observations have shown even better long-term radial velocity stability of 0.37 m~s$^{-1}$ over a period of 3.5 years after activity mitigation \citep{ford_2024}.

We use NEID DRP v1.4 observations from February 1, 2021 to September 8, 2025, which extends beyond the range used in \citet{ford_2024} using NEID DRP v1.3. 
Appendix \ref{Comparisons between NEID v1.3 and v1.4 Data} compares the RV performance of NEID DRP v1.3 and v1.4, since the transition between these two versions introduced changes to the algorithms used to derive the wavelength calibrations after the LFC was upgraded with a new photonic crystal fiber (PCF)\footnote{\url{https://neid.ipac.caltech.edu/docs/NEID-DRP/versions.html}}.

With the longer timespan of observations available using NEID DRP v1.4, we split the range of observations into three runs, described in Table \ref{tab:observation_vetting}.
The Contreras Fire caused a pause of NEID solar observations, with the final pre-fire data taken on June 13, 2022. 
NEID resumed limited operations on November 12, 2022 while cooling back down with high-precision night observations restarting on November 23, 2022 and full operations resuming December 1, 2022.
We exclude data that was collected between the fire and December 1, 2022 and define data taken during February 1, 2021 $–$ June 13, 2022 as ``run 1”. The LFC PCF was upgraded during the week of August 24, 2024, so we define data taken during December 1, 2022 $–$ June 30, 2024 as ``run 2”. Finally, we define data taken during February 19, 2025 $–$ June 21, 2025 as ``run 3”. 
We choose to end ``run 3” on June 21, 2025 because the PCF was replaced and this changed the wavelength solution for some blue orders, causing spurious RVs for lines in these orders.\footnote{These date ranges are more restrictive than eras listed in \url{https://neid.ipac.caltech.edu/docs/NEID-DRP/rveras.html}, reflecting the availability of high-quality solar observations. We anticipate that subsequent versions of the NEID pipeline will likely enable the use of more data.}  

\begin{table*}[htb]
\centering
\caption{Summary of NEID solar observation runs used in this work, including the number of days and exposures per run and the corresponding thresholds for the mean pyrheliometer flux and mean exposure meter flux. 
These thresholds were chosen to filter approximately the same fraction of observations in each run.}
\label{tab:observation_vetting}
\begin{tabular}{||c c c c c c c||} 
 \hline
 Run & Start & End & $\#$ Days & $\#$ Exposures & pyr$\_$flux$_\text{threshold}$ (W m$^{-2}$) & exp$\_$meter$_\text{threshold}$ (ADU) \\ [0.5ex] 
 \hline\hline
1 & 2021/02/01 & 2022/06/13 & 149 & 32,876 & $10^{2.95}$ & $10^5$ \\[1ex] 
2 & 2022/12/01 & 2024/06/30 & 170 & 39,794 & $10^{2.943}$ & $10^{4.9}$ \\ [1ex] 
3 & 2025/02/19 & 2025/06/21 & 64 & 15,172 & $10^{2.93}$ & $10^{4.8}$ \\ [1ex] 
 \hline
\end{tabular}
\end{table*}

Given the success of \citet{ford_2024} in analyzing NEID solar observations and modeling stellar activity, we adopt a similar framework for filtering and vetting data, as described in Appendix \ref{Vetting of NEID Observations}.
Since the conditions of the instrument varied between runs, we select flux thresholds for each run such that they filter approximately the same proportion of exposures among the runs, and these thresholds are listed in Table \ref{tab:observation_vetting}.  

\subsection{Constructing Daily-Averaged Spectra via Gaussian Process} \label{Constructing Daily-Averaged Spectra via Gaussian Process} 

To increase the SNR of our observations, average over short-term stellar activity signals such as p-modes (with a period of $\sim$5 mins for the Sun, \citealp{kjeldsen_2008}) and granulation ($\sim$few hours for the Sun, \citealp{meunier_2015}), and to reduce computational time of our LBL fitting, we average exposures within a day into daily-averaged spectra. 
We use the Penn State Research Pipeline (PSRP), which includes multiple software packages in the RvSpectML organization\footnote{\url{https://rvspectml.github.io/RvSpectML.jl/stable/}} (see Section 3 of \citealp{ford_2024}). 
We begin from order-extracted, blaze-normalized spectra for each exposure in a day from Section \ref{NEID Solar Data} and use a Matern-$\frac{5}{2}$ Gaussian Process (GP) conditioned on all accepted exposures in a day to evaluate the posterior mean flux and the posterior variance along a standard grid of wavelengths with the same number of pixels as the original spectrum. 
We choose a GP smoothing factor\footnote{This factor scales the kernel variance and length scale of the GP} of 1, which is RvSpectML's default value and works well visually. 
We process all physical \'echelle orders of the spectra (55 to 173) even though orders $\sim 55 - 60$ and $\sim 163 - 173$ have small SNR. 
Once we have our daily-averaged spectra, we normalize the flux in each day and order by the 95$^{\text{th}}$ percentile of that order's flux. 

\subsection{Line List} \label{line list}

We use RvSpectML's default line list\footnote{\url{https://github.com/RvSpectML/RvLineList.jl}} (Wise et al. in prep.). 
Appendix \ref{Line List Construction} explains how this line list was constructed. 
We treat the same line but in different \'echelle orders as different lines. 
Therefore, even though the line list has 1,579 lines with distinct wavelengths, there are a total of 3,830 line/order pairs in our analysis. 

We note that this line list contains lines that are outside the NEID free spectral range (FSR)\footnote{\url{https://neid.ipac.caltech.edu/docs/NEID-DRP/masterfiles.html\#echellogram-definition}} in a given order. 
For comparison, the NEID DRP excludes such lines when constructing order-by-order masks and computing RVs. 
In this work, we keep these lines to empirically assess how their inferred RV behavior depends on detector geometry such as pixel and order position, and how much RV information can be extracted for lines outside the nominal boundaries used by the DRP. 

\subsection{Line Fitting} \label{Line Fitting}

For each line in our line list, we construct a 15-pixel window around the line center (see Appendix \ref{Window Type} for an explanation of this choice). 
We then use a least-squares approach to fit a Gaussian with a continuum component to the line. 
This gives us an approximate time-dependent central wavelength of the line. 
The procedure of fitting a Gaussian model to each line is further explained in Appendix \ref{Gaussian Model for Line Fitting}. 

For each line $l$ and time $t$, we calculate the Doppler RV using
\begin{equation}
\text{RV}_{l, \text{t}} = c \left( \frac{\lambda_{l, \text{t}} - \lambda_{\text{rest}, l}}{\lambda_{\text{rest}, l}} \right),
\label{EQ:doppler}
\end{equation}
where $\lambda_{l, \text{t}}$ is the observed wavelength of the line $l$ at time $t$ (taken as the central wavelength of a Gaussian fit, as explained in Section \ref{Gaussian Model for Line Fitting}), $\lambda_{\text{rest}, l}$ is the rest wavelength of line $l$ which is taken from the Vienna Atomic Line Database (VALD, \citealp{ryabchikova_2015}), and $c$ is the speed of light. 

We also estimate the corresponding uncertainty $\sigma_{l,t}$ for each $\text{RV}_{l, \text{t}}$ by simulating 100 independent and identically distributed realizations of each line by injecting it with photon noise, refitting the Gaussian model, and recomputing the RV (see Appendix \ref{Gaussian Model for Line Fitting}). 
In addition to uncertainties for the RV, we also similarly compute uncertainties for each shape parameter. 

We subtract the mean RV of each line within each run to remove run-to-run instrumental offsets.\footnote{see the NEID documentation: \newline \url{https://neid.ipac.caltech.edu/docs/NEID-DRP/rveras.html}}  
This also removes the constant offset from the mean barycentric correction of the run, the gravitational redshift, and the mean convective blueshift (``Third Signature" of granulation, \citealp{gray_2005}).

\subsection{Line Vetting \label{Line Vetting}}

After computing RVs (and corresponding uncertainties) for all 3,830 lines and 383 days, we implement a vetting procedure to reject lines whose RV measurements significantly deviate from the RVs of all lines and whose discrepancies cannot be explained by photon noise. 
This is because certain groups of lines exhibit large RV deviations from the rest of the lines. 
This is most notable during run 3 and likely due to the updated algorithms to derive wavelength calibrations after the LFC was upgraded with a new PCF at the end of run 2, which caused some lines to show spurious RVs. 
There are also some lines in the reddest orders whose RVs deviate significantly from the RVs of all other lines throughout all days. 

We first compute line-weighted RVs for each day. 
Let RV$_{l,t}$ be the radial velocity measured for spectral line $l$ at time $t$ with associated uncertainty $\sigma_{l,t}$ estimated from bootstrapped photon-noise simulations (see Appendix \ref{Gaussian Model for Line Fitting}). 
We compute a weighted RV time series by first assigning each line a scalar weight based on its long-term RV precision over all days, given as
\begin{equation}
W_l = \mathrm{percentile}\left( \left[ \frac{1}{\sigma_{l,0}^2}, \frac{1}{\sigma_{l,1}^2}, \dots, \frac{1}{\sigma_{l,T}^2} \right], 10 \right).
\end{equation}
Using a low percentile down-weights lines that are usually precise but occasionally exhibit large uncertainties.
We then normalize the weights:
\begin{equation}
\sum_{l=1}^{N_\ell} W_l = 1.
\end{equation}
The weighted-mean RV time series is then 
\begin{equation}
    \mathrm{RV}^{\mathrm{wt}}_t
    = \sum_{l=1}^{N_\ell} W_l \, \text{RV}_{l,t},
    \label{EQ:weighted_rv_timeseries}
\end{equation}
with corresponding uncertainty
\begin{equation}
    \sigma^{\mathrm{wt}}_t
    = \sqrt{ \sum_{l=1}^{N_\ell} W_l^2 \, \sigma_{l,t}^2 } .
    \label{EQ:weighted_rv_uncertainty}
\end{equation}

\begin{figure}
    \centering
	\includegraphics[width=8 cm]{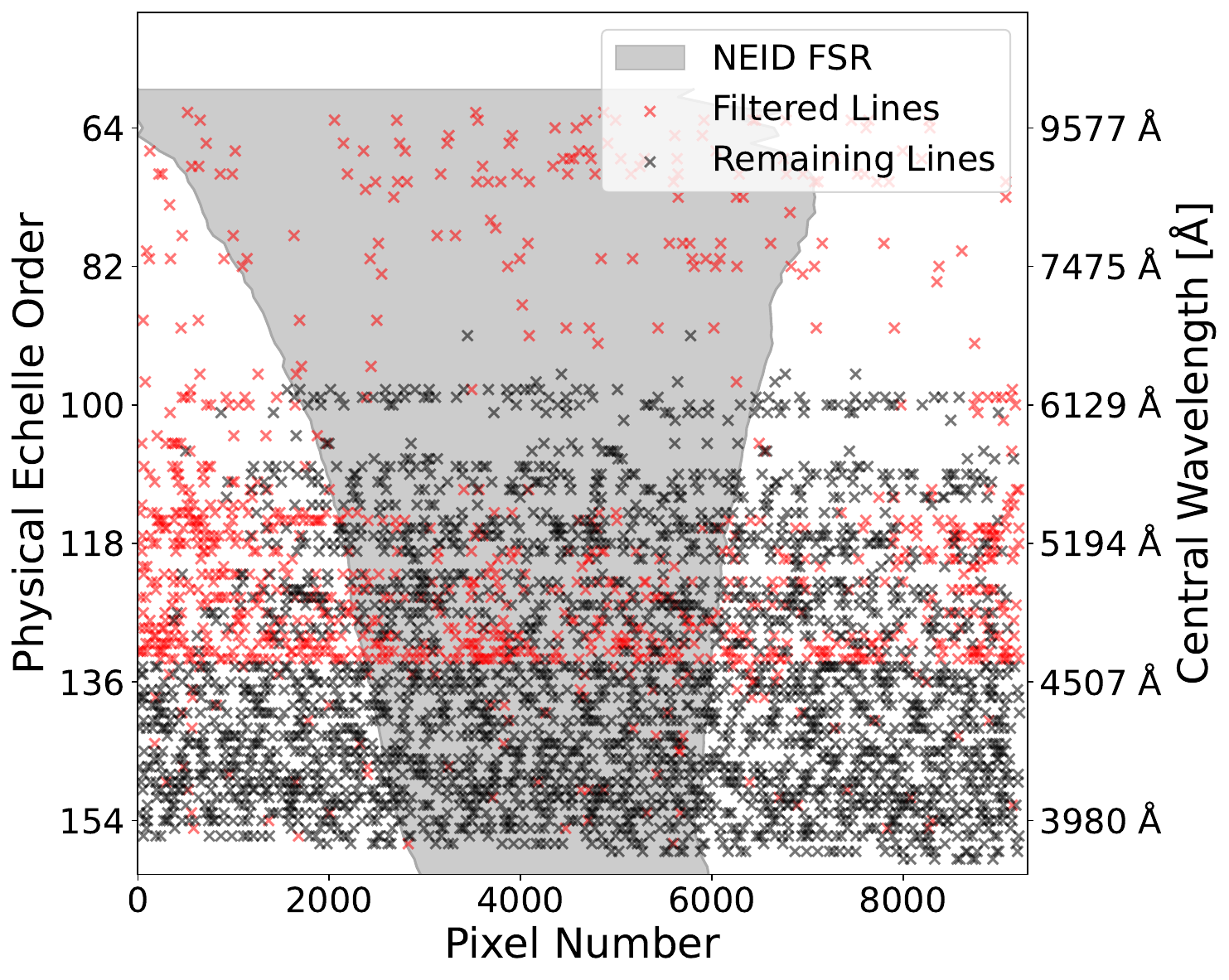}
    \caption{The filtered (red) and remaining (black) lines on the detector after removing outlier lines that deviate from the weighted time average, described in Section \ref{Line Vetting} and further in Appendix \ref{Vetting Lines that deviate from Weighted Time Average}. 
    The gray shade shows an approximate projection of the NEID FSR onto the extracted orders. 
    All lines redward of physical \'echelle order 85 are removed which is likely due to telluric contamination, and a significant fraction of lines between physical \'echelle orders $114-134$ are removed due to the LFC upgrade at the end of run 2. }
    \label{fig:detector}
\end{figure}

Appendix \ref{Vetting Lines that deviate from Weighted Time Average} describes the procedure to remove lines that deviate significantly from the line-weighted RV time series using two separate filtering passes. 
The locations of filtered and remaining lines on the detector after the second filtering pass are shown in Figure~\ref{fig:detector}. 
All lines redward of physical \'echelle order 85 are removed which is likely due to telluric contamination. Also, a significant fraction of lines between physical \'echelle orders $114-134$ are removed due to the LFC upgrade at the end of run 2.  
These are the bluest LFC orders, which have the worst flux variability issues. 
Many of these may become useful with v1.5 of the NEID data reduction pipeline.  
The lines that remain redward of physical \'echelle order 134 tend to be in the center of the detector, where the wavelength calibration is best constrained due to higher SNR near the blaze peak. 

The final vetted line list of $2,809$ addresses the known instrumental issues introduced by the PCF upgrade as well as the poorly behaved redder orders that are contaminated by tellurics.
We therefore adopt this curated line list for all subsequent analyses in this work.
We release a dataset of LBL RVs (and corresponding uncertainties) across all 383 days of observations on Zenodo \citep{zenodo_pandey_2026}. 

\subsection{Autocorrelation and Activity Diagnostics of LBL RVs \label{Autocorrelation and Activity Diagnostics of LBL RVs}}

We find that computing a line-weighted RV time series gives an RV RMS of $\approx 2.012~\mathrm{m\,s^{-1}}$ using all lines.  
This is similar to an RV RMS of $\approx 2.129~\mathrm{m\,s^{-1}}$ using the CCF approach based on \citet{ford_2024} (see Appendix \ref{LBL vs CCF RVs}).

We also quantify quasi-periodic variability in an RV time series by computing the autocorrelation of each RV signal at time lags of 28 and 365 days, which are roughly the solar rotation period and the annual observing cycle. These lags allow us to assess how rotationally modulated and long-term seasonal signals affect RV variability.

The following modified autocorrelation function (ACF) quantifies how similar the RV measurements are when separated by a lag of $\tau$:
\begin{comment}
\begin{equation}
\mathrm{ACF}(\tau) = 
\frac{1}{N(\tau)\,\sigma_{\mathrm{RV}}^2}
\sum_{\substack{i < j \\ |(t_j - t_i) - \tau| \leq \Delta\tau}}
\bigl(\mathrm{RV}_i - \overline{\mathrm{RV}}\bigr)
\bigl(\mathrm{RV}_j - \overline{\mathrm{RV}}\bigr),
\label{EQ:ACF}
\end{equation}
\end{comment}
\begin{equation}
\begin{aligned}
&\mathrm{ACF}(\tau) = \frac{1}{N(\tau)\,\sigma_{\mathrm{RV}}^2} \\[6pt]
&\times \sum_{\substack{i < j \\ |(t_j - t_i) - \tau| \leq \Delta\tau}}
 \bigl(\mathrm{RV}_i - \overline{\mathrm{RV}}\bigr)
 \bigl(\mathrm{RV}_j - \overline{\mathrm{RV}}\bigr),
\end{aligned}
\label{EQ:ACF}
\end{equation}
where $\mathrm{RV}_i$ is the radial velocity measurement observed at time $t_i$, $\overline{\mathrm{RV}}$ is the mean of all RV measurements, $\sigma_{\mathrm{RV}}^2$ is the variance of the RV series, $\tau$ is the target time lag at which the autocorrelation is evaluated, $\Delta\tau = 0.1 \tau$ is the lag tolerance, and $N(\tau)$ is the number of pairs $(i, j)$ such that $|(t_j - t_i) - \tau| \leq \Delta\tau$ \citep{edelson_1988}. 
This modified autocorrelation function accounts for uneven time sampling in RV data by allowing a small deviation around the target lag $\tau$. 
The allowed deviation $\Delta\tau = 0.1 \tau$ ensures that nearly matching observation pairs contribute to the autocorrelation. 
Since the correlation is computed over a finite lag bin, the resulting statistic is not strictly guaranteed to be bounded within $[-1, 1]$, although all values reported in this work fall within that range.
Equation \eqref{EQ:ACF} shows that a positive $\mathrm{ACF}$ evaluated at a lag of $\tau$ indicates that RVs tend to vary coherently on timescales near $\tau$, whereas negative values indicate anti-correlation (i.e., RVs oscillate with that period). 

For a given RV time series, we define ACF30 as the ACF evaluated at a time lag of 28 days, which is roughly the solar rotation period. 
We also define ACF365 as the ACF evaluated at a time lag of 365 days which is the annual observing cycle. 
For the LBL RVs, the ACF30 is $\approx$ 0.622 and ACF365 $\approx-0.08$, which are similar to ACF30 $\approx$ 0.648 and ACF365 $\approx-0.116$ for the CCF RVs (see Appendix \ref{LBL vs CCF RVs}). 

In addition to the RV RMS, ACF30, and ACF365, we also compute $\rho$(RV, Ca\,\textsc{ii}\,HK), which is the Spearman correlation coefficient between a given RV time series and the Ca\,\textsc{ii}\,HK activity index (a common proxy for chromospheric activity). 
For a given day, we use the median Ca\,\textsc{ii}\,HK activity index values from the NEID DRP\footnote{\url{https://neid.ipac.caltech.edu/docs/NEID-DRP/algorithms.html\#stellar-activity-info}} for each daily observation. 
We find $\rho$(RV, Ca\,\textsc{ii}\,HK)$\approx0.172$ for the LBL RVs which is slightly lower than $\rho$(RV, Ca\,\textsc{ii}\,HK)$\approx0.208$ for the CCF RVs.
However, both RV time series show weak correlation with Ca\,\textsc{ii}\,HK. 

\section{Benchmark Line Selection Algorithms \label{Benchmark Line Selection Algorithms}}

Section \ref{Autocorrelation and Activity Diagnostics of LBL RVs} showed that using RV information from all 2,809 lines achieved an RV RMS of $\approx 2.012~\mathrm{m\,s^{-1}}$. 
However, not all spectral lines contribute equally to RV precision. 
The full line list includes lines with low SNR, shallow lines, lines affected telluric contamination, and features located near detector edges where instrumental systematics are more pronounced. 
It can therefore be advantageous to identify subsets of spectral lines that are less sensitive to stellar activity, instrumental effects, and telluric contamination. 
In this section, we introduce several benchmark methods for selecting such subsets of lines. 
These approaches are a useful comparison against an iterative algorithm which we will later introduce in Section~\ref{Filtering Lines for Accurate RV Exoplanet Surveys (FLARES)}.

\begin{figure*}
    \centering
	\includegraphics[width=18 cm]{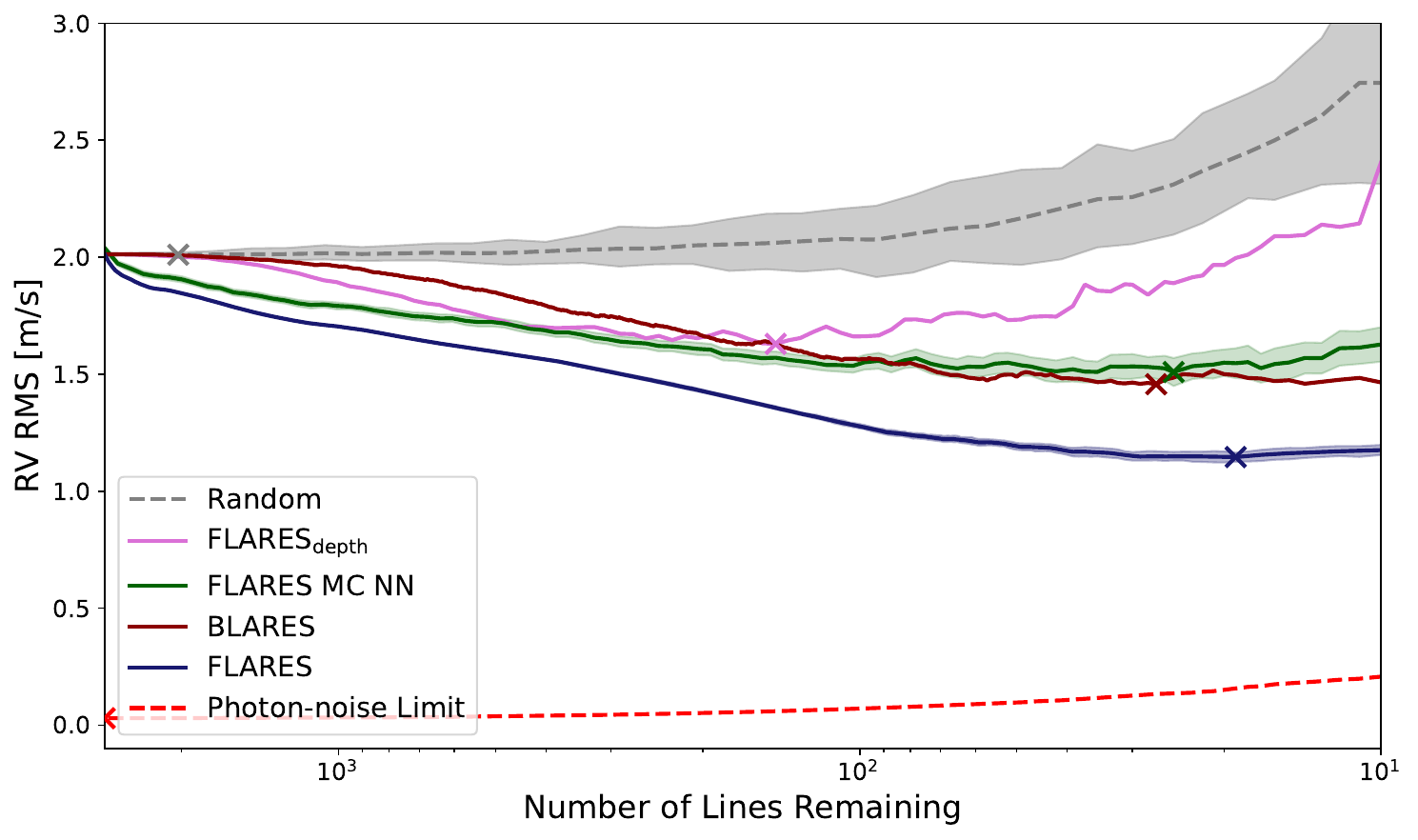}
    \caption{Evolution of RV RMS as a function of the number of spectral lines retained for Random (dashed gray; see Section \ref{Random}), BLARES (solid dark-red; see Section \ref{BLARES}), FLARES$_{\textrm{depth}}$ (solid pink; see Section \ref{1D Removal based on Spectral Line Depth}), FLARES (solid dark-blue; see Section \ref{Filtering Lines for Accurate RV Exoplanet Surveys (FLARES)}), and FLARES MC NN (solid dark-green; see Section \ref{Constructing Comparison Line Lists}). 
    Also shown is the expected photon-noise limit (dashed red; see Section \ref{Photon-noise Limit}).
    Shaded regions for Random, FLARES, and FLARES MC NN depict the $1\sigma$ dispersion across 100 bootstrap simulations. 
    Colored "X" markers show how many lines are retained when the RV RMS is minimized for each method (see Table \ref{tab:minimum_rv_rms_comparison}). 
    FLARES and BLARES consistently have a lower RV RMS than Random, and FLARES shows $\sim0.3~\mathrm{m\,s^{-1}}$ lower RV RMS than BLARES. 
    However, the expected photon-noise limit is much lower than both BLARES and FLARES.
    FLARES MC NN outperforms Random but not FLARES, suggesting that matching the properties of the FLARES-selected lines improves RV precision but does not fully reproduce the performance of FLARES (see discussion in Section \ref{Constructing Comparison Line Lists}).
    }
    \label{fig:rv_rms_all_methods}
\end{figure*}

Figure \ref{fig:rv_rms_all_methods} shows the evolution of the RV RMS as a function of the number of spectral lines in the subset using several benchmark algorithms discussed in this section as well as other methods discussed later in this work. 
Table \ref{tab:minimum_rv_rms_comparison} shows the minimum RV RMS achieved using these methods, the number of spectral lines retained at the minimum RV RMS for each method, and the section where each method is introduced in this work. 

\begin{table}[hbt!]
\centering
\caption{Minimum RV RMS achieved by each method in this work, along with the number of spectral lines retained at the minimum.}
\label{tab:minimum_rv_rms_comparison}
\begin{tabular}{lccc}
\hline
Method & Section & Min. RMS [m\,s$^{-1}$] & \# Lines \\
\hline
Random              & \S\ref{Random} & 2.011 & 1956 \\
FLARES$_{\textrm{depth}}$
                    & \S\ref{1D Removal based on Spectral Line Depth} & 1.630 & $\approx$150 \\
FLARES MC NN        & \S\ref{Constructing Comparison Line Lists} & 1.510 & $\approx$25 \\
BLARES              & \S\ref{BLARES} & 1.457 & 27 \\
FLARES MC           & \S\ref{Monte Carlo Simulations of FLARES} & 1.150 & $\approx$25 \\
FLARES              & \S\ref{Filtering Lines for Accurate RV Exoplanet Surveys (FLARES)} & 1.122 & 24 \\
Photon-noise Limit  & \S\ref{Photon-noise Limit} & $\sim0.05$ & 2,809 \\
\hline
\end{tabular}
\end{table}

\subsection{Photon-noise Limit \label{Photon-noise Limit}}
The photon-noise limit provides a fundamental lower bound on the RV precision achievable from a given set of spectral lines under the assumption that all noise is purely due to photon statistics. 
For methods introduced later in this work, we can asses how far each method is from this photon-noise limit.

We compute the expected photon-noise limited RV precision by combining the inverse-variance photon-noise of every line's RV uncertainties. 
Figure \ref{fig:rv_rms_all_methods} shows that when using all lines, the photon-noise limit approaches $\sim 0.03-0.04~\mathrm{m\,s^{-1}}$ for daily-averaged NEID solar observations. 
This is slightly lower than the $\sim 0.045~\mathrm{m\,s^{-1}}$ photon-noise limit reported in \citet{ford_2024}, although that value was computed using only one hour of integration time.
This is consistent with the $\sim0.02$--$0.07~\mathrm{m\,s^{-1}}$ range reported by \citet{ford_2024} for their daily binned spectra, where the photon-noise limit depends on the number of observations included.
As the subset of lines becomes smaller, the photon-noise limit approaches $\sim0.2~\mathrm{m\,s^{-1}}$ for the best $20-30$ lines.

\subsection{Random \label{Random}}
A baseline model for line selection that preserves no information about line quality of physical properties is computing the RV RMS on $N$ randomly chosen lines from the full sample. 
We refer to this as Random for the rest of this work.
To quantify the scatter in the RV RMS due to the stochastic nature of Random, we repeat this 100 times for each $N$ and compute the 1$\sigma$ dispersion of outcomes. 
Figure \ref{fig:rv_rms_all_methods} shows that the RV RMS of Random generally increases and is less precise with fewer lines, and increases drastically to $> 2.5$ m~s$^{-1}$ when there are fewer than 100 lines. 
The lowest RV RMS achieved using Random is $\approx 2.011~\mathrm{m\,s^{-1}}$ using $\approx2000$. 

\subsection{Building Linelists for Accurate RV Exoplanet Surveys (BLARES) \label{BLARES}} 

If only the RV time series of each line is available, a natural first step in constructing optimal line subsets is to rank all lines by their LBL RV RMS. 
To select a subset of the best $N$ lines, one can then simply choose the $N$ lines with the lowest RV RMS. 
We refer to this as Building Linelists for Accurate RV Exoplanet Surveys (BLARES).
We release a dataset of LBL rankings based on when lines were removed by BLARES on Zenodo \citep{zenodo_pandey_2026}. 

Figure \ref{fig:rv_rms_all_methods} shows that BLARES achieves a minimum RV RMS of $1.457~\mathrm{m\,s^{-1}}$ using only 27 lines, which is significantly lower than Random.
Since solar RVs have been corrected for known planetary signals, a lower RV RMS indicates that the selected lines are less sensitive to sources of RV variability, including stellar activity, instrumental systematics, and telluric contamination.
This suggests that building line lists purely using LBL RV RMS as a way to rank lines is a reasonable approach in identifying optimal line subsets.  
However, the photon-noise limit approaches $\sim0.2~\mathrm{m\,s^{-1}}$ for the best $20-30$ lines which suggests that most of the RV signal from BLARES is still arising from some combination of stellar variability, instrumental systematics, and telluric contamination.

\subsection{1D Binning based on Spectral Line Depth \label{1D Binning based on Spectral Line Depth}}

Although BLARES identifies stable lines using only their observed RV scatter, it does not provide direct insight into the physical properties that make certain lines more stable than others. 
Previous studies have shown that spectral line depth can be an important indicator of activity sensitivity \citep{meunier_2017, cretignier_2020, al_moulla_2023}.
Spectral line depth is physically linked to line formation height in the photosphere \citep{gray_2005}. 
Shallower lines tend to form deeper in the atmosphere where convective blueshift is stronger, making them more sensitive to magnetic activity and its inhibition of convective blueshift as compared to deeper lines \citep{meunier_2017, cretignier_2020, al_moulla_2023}. 
We therefore investigate how RV statistics depend on line depth and whether physically motivated line selection by binning can achieve similar RV RMS as BLARES. 

Our first approach is to bin lines by line depth and find subsets of the deepest lines to mitigate the effects of stellar activity.
We use line depths predicted by VALD \citep{ryabchikova_2015}.\footnote{Although line depths can be measured directly from the spectra (see Section \ref{Gaussian Model for Line Fitting}), we adopt VALD values to avoid introducing run-dependent variations across the time series. We also verified that VALD depths are strongly correlated with depths measured from spectra.}
We sort the 2,809 spectral lines by depth and divide them into 10 bins\footnote{The choice of 10 bins is arbitrary, but the results are robust and stable when also considering 4 and 20 bins.}  such that each bin contains $\approx 281$ lines.  
We also bootstrap 1,000 randomly generated line lists of the same size as each bin to illustrate the expected behavior of typical random clusters. 

\begin{figure}[hbt!]
    \centering
	\includegraphics[width= 8.5 cm]{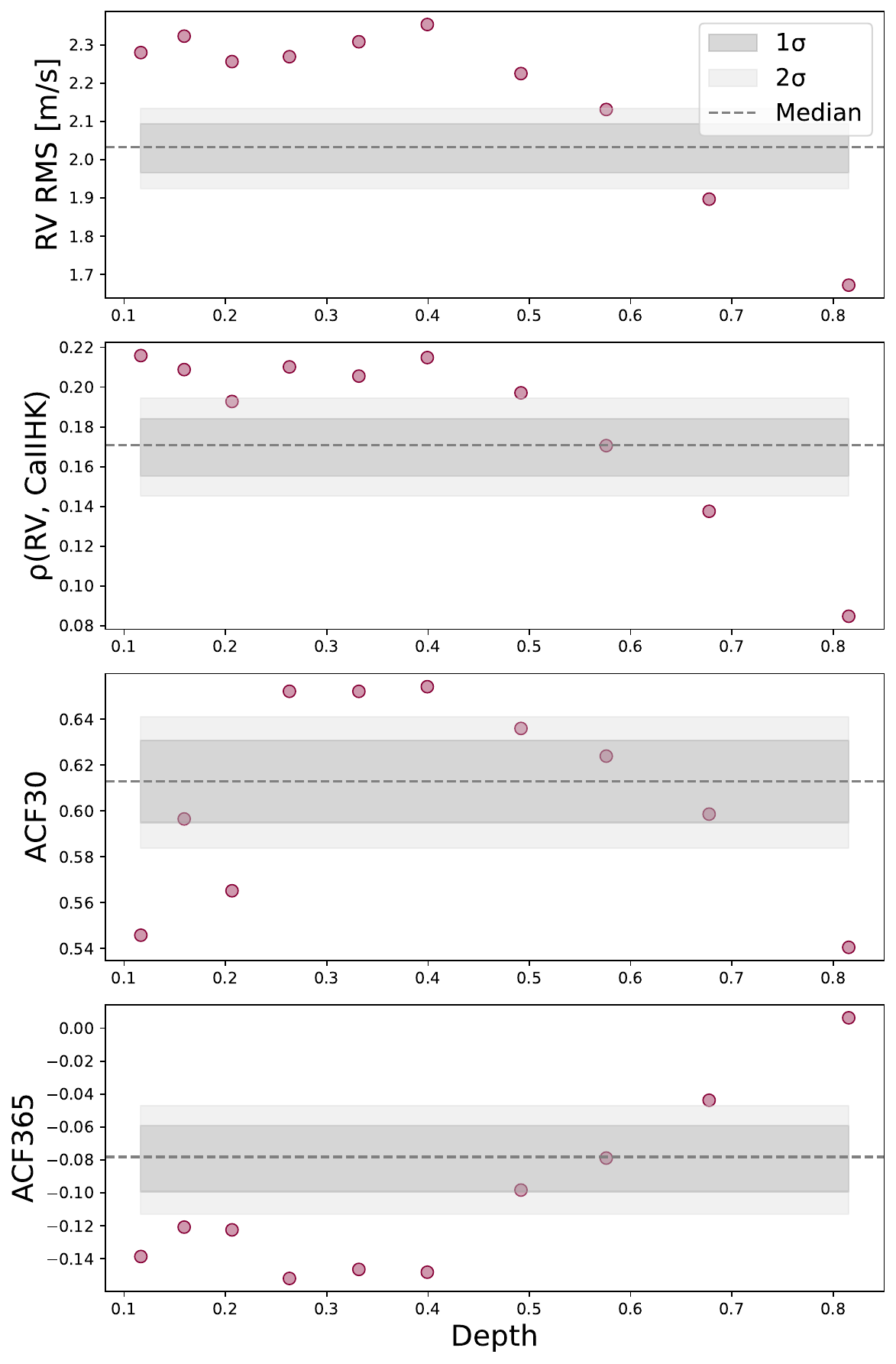}
    \caption{The dependence of RV properties on spectral line depth. 
    All $2,809$ lines are sorted by depth and divided into 10 bins such that each bin contains $\approx 281$ lines (see Section \ref{1D Binning based on Spectral Line Depth}). 
    For comparison, each panel shows the median (dashed black) and the $1\sigma$ (dark gray) and $2\sigma$ (light gray) dispersion across 1,000 bootstrap simulations of randomly generated line lists with the same size as each bin to illustrate the expected behavior of typical random clusters. 
    \textit{Top:} RV RMS for binned subsets of spectral lines as a function of line depth. 
    The RV RMS is $\approx 2.3~\mathrm{m\,s^{-1}}$ for the bins with the shallowest lines, but decreases to $\approx 1.7~\mathrm{m\,s^{-1}}$ for the bins with the deepest lines. 
    \textit{Top-middle:} the Spearman correlation coefficient between the RVs in each depth bin and the Ca\,\textsc{ii}\,HK activity index. 
    The shallowest bins exhibit the largest correlation ($\rho \approx 0.21$) between their RVs and Ca\,\textsc{ii}\,HK, whereas deepest bins exhibit weaker correlations ($\rho \approx 0.08$). 
    \textit{Bottom-middle:} the ACF30 for RVs in each depth bin. 
    Both the shallowest and deepest bins exhibit a low periodicity (ACF30 $\approx 0.54$), whereas intermediate-depth lines have the largest periodicity (ACF30 $\approx 0.65$). 
    \textit{Bottom:} the ACF365 for RVs in each depth bin. The shallowest-depth bins show the largest deviation from 0 (ACF365 $\approx -0.15$) whereas the deepest bins show no such periodic signal (ACF365 $\approx 0$). 
    }
    \label{fig:1D_binning_Depth}
\end{figure}

Figure \ref{fig:1D_binning_Depth} shows the RV RMS, $\rho$(RV, Ca\,\textsc{ii}\,HK), ACF30, and ACF365 for every binned subset. 
The shallowest lines have a large RV RMS of $\approx 2.3~\mathrm{m\,s^{-1}}$, along with more correlation with Ca\,\textsc{ii}\,HK ($\rho$(RV, Ca\,\textsc{ii}\,HK) $\approx$ 0.22). 
In contrast, deeper lines have a smaller RV RMS of $\approx 1.7~\mathrm{m\,s^{-1}}$ and show less correlation with Ca\,\textsc{ii}\,HK ($\rho$(RV, Ca\,\textsc{ii}\,HK) $\approx$ 0.08). 
These results are consistent with the differential inhibition of convective blueshift due to magnetic fields between shallow and deep lines. 
The finding that the activity-induced RV variability is inversely proportional to line depth has also been shown on $\alpha$ Cen B using HARPS-N observations \citep{cretignier_2020}. 

Figure \ref{fig:1D_binning_Depth} also shows that the ACF30 generally decreases moving from shallow lines (depths between $0.25 \leq d \leq 0.45$) with an ACF30 of $\approx$ 0.65 to deeper bins (depths $d \geq 0.8$) with an ACF30 of $\approx$ 0.55. 
The shallowest bins also show a reduced ACF30 $\approx$ 0.55.
Although shallow lines are often expected to show stronger rotational modulation due to enhanced activity sensitivity and the differential suppression of convective blueshift effect as compared to deep lines, we do not observe a simple monotonic dependence of ACF30 on line depth. 
Deep lines likely show reduced ACF30 due to the lower sensitivity to activity, whereas the shallow lines likely show reduced ACF30 due to reduced SNR and lower RV information content which produces a noisier time series and dilutes the ACF30 signal. 

Finally, the ACF365 approaches $\approx$ 0 for the deepest bins, whereas the shallowest bins show a slight negative power of ACF365 $\approx -0.14$. 
This could be due to instrumental effects or telluric contamination and the seasonal effects of changing airmass which more strongly affects the shallowest lines. 

These results suggest that line depth is a useful physically motivated metric for identifying stable spectral lines, and this motivates the line-selection approach explored in the following section.

\subsection{1D Removal based on Spectral Line Depth \label{1D Removal based on Spectral Line Depth}}

The results presented in Section \ref{1D Binning based on Spectral Line Depth} suggest that binning purely based on spectral depth and only focusing on the deepest lines that are least affected by stellar activity and contain the most RV information have a lower RV RMS than Random but still do not perform as well as BLARES. 
We therefore combine these ideas by developing an iterative removal scheme similar to BLARES to generate line lists of a given size, but guided by line depth rather than LBL RV RMS. 
The results of constructing such line lists using only the depth will provide a useful comparison for FLARES (see Section \ref{Filtering Lines for Accurate RV Exoplanet Surveys (FLARES)}.)

First, we sort all 2,809 lines by depth and divide them into 100 bins such that each bin contains $\approx 28$ lines with similar depths. 
We then iteratively remove the bin containing the shallowest lines and recompute the same RV statistics shown in Figure \ref{fig:1D_binning_Depth} using the remaining lines.
For the rest of this section, we refer to this method as FLARES$_{\textrm{depth}}$, as it is similar to the iterative filtering algorithm which will be introduced in Section \ref{Filtering Lines for Accurate RV Exoplanet Surveys (FLARES)}, but only using depth as an input metric. 

For comparison, we again bootstrap 10,000 random line lists, each with size drawn from a log-uniform prior between 10 and 2,809 lines, and compute the same RV statistics. 
These bootstrap samples illustrate the expected behavior of typical random subsets of lines.

\begin{figure}[hbt!]
    \centering
	\includegraphics[width= 8.5 cm]{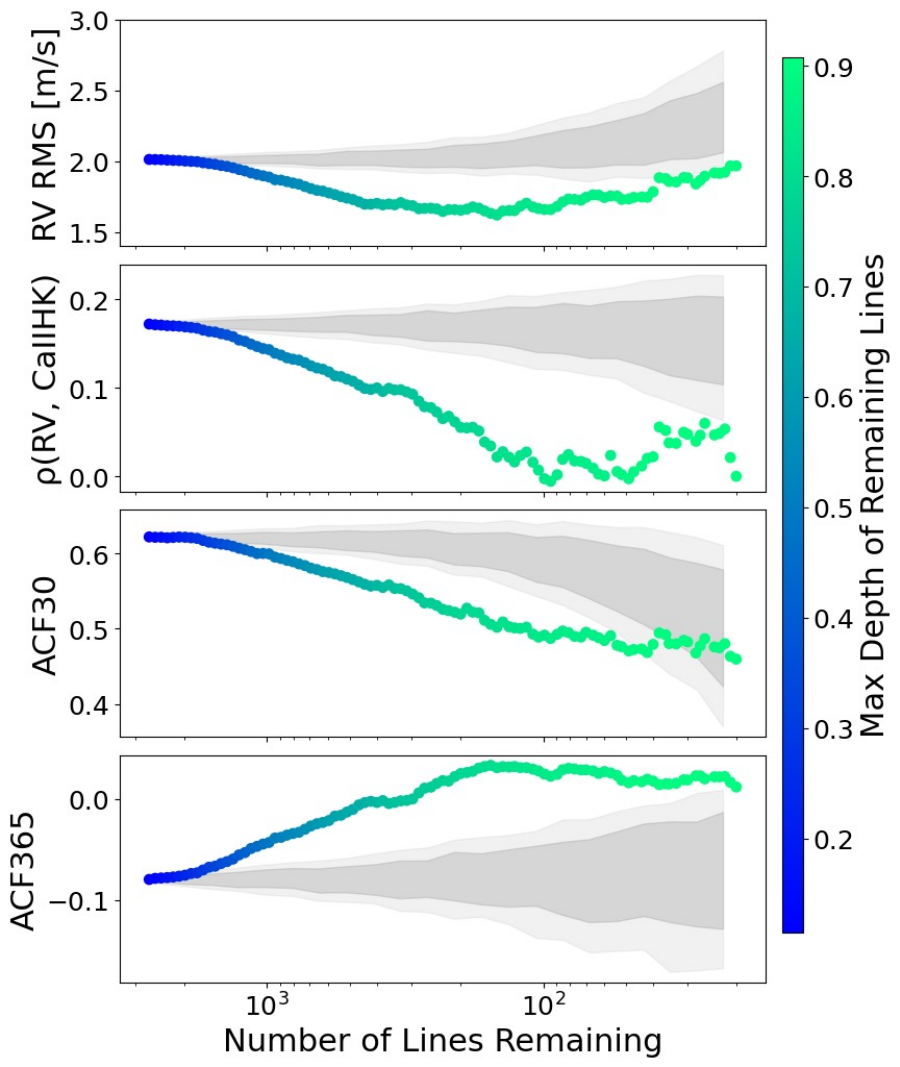}
    \caption{The evolution of RV properties as the shallowest lines are iteratively removed using FLARES$_\textrm{depth}$ (see Section \ref{1D Removal based on Spectral Line Depth}). 
    For comparison, each panel includes the $1\sigma$ (gray) and $2\sigma$ (light gray) dispersion across 10,000 bootstrap simulations of randomly generated line lists, with sizes drawn from a log-uniform prior spanning 10 to 2,809 lines, to illustrate the expected behavior of typical random clusters. 
    \textit{Top:} RV RMS as a function of the number of lines remaining. 
    The RV RMS decreases as the shallowest lines are removed and fall below the distribution of random subsamples. 
    The RV RMS reaches $\approx 1.6~\mathrm{m\,s^{-1}}$ for the deepest (d $\geq 0.7$) $\approx 200$ lines. 
    \textit{Top-middle:} the Spearman correlation coefficient between the RVs and the Ca\,\textsc{ii}\,HK activity index as a function of the number of lines remaining. 
    The correlation decreases as shallower lines are removed, reaching $\rho \approx 0.05$ for the deepest ($d \geq 0.7$) $\approx 200$ lines. 
    \textit{Bottom-middle:} the ACF30 as a function of the number of lines remaining. 
    As shallower lines are removed, the ACF30 decreases and falls below the distribution of random subsamples. 
    The ACF30 is $\approx 0.5$ for the deepest (d $\geq 0.7$) $\approx 200$ lines. 
    \textit{Bottom:} the ACF365 as a function of the number of lines remaining. 
    As shallower lines are removed, the ACF365 approaches 0 for the deepest (d $\geq 0.7$) $\approx 200$ lines. 
    }
    \label{fig:1D_removal_Depth}
\end{figure}

Figure \ref{fig:1D_removal_Depth} shows that as shallower lines are removed, the RV RMS of the remaining deep lines fall below the distribution of Random. 
In addition, the deepest (d $\geq$ 0.7) $\approx$ 150 lines achieve an RV RMS of 1.63 m~s$^{-1}$ which is comparable to the RV RMS achieved by BLARES (see Figure \ref{fig:rv_rms_all_methods} and Table \ref{tab:minimum_rv_rms_comparison}).
This suggests that it is possible to use line properties other than LBL RV RMS to find optimal line subsets.  

The RVs of all lines show $\rho$(RV, Ca\,\textsc{ii}\,HK) $\approx 0.17$. 
As the shallowest lines are removed using FLARES$_{\textrm{depth}}$, the remaining deep lines show less correlation as compared to Random and have $\rho$(RV, Ca\,\textsc{ii}\,HK) $\approx$ 0.05 for the deepest (d $\geq$ 0.7) $\approx$ 200 lines.

Figure \ref{fig:1D_removal_Depth} also shows that all lines have a large ACF30 $\approx$ 0.6. 
As the shallowest lines are removed using FLARES$_{\textrm{depth}}$, the remaining deep lines show less power at ACF30 as compared to Random, and the deepest (d $\geq$ 0.7) $\approx$ 200 lines have an ACF30 of $\approx 0.5$. 

Finally, all lines show ACF365 $\approx$ $-$0.08. 
As the shallowest lines are removed using FLARES$_{\textrm{depth}}$, the remaining deep lines show less power at ACF365 as compared to Random and the deepest (d $\geq$ 0.7) $\approx$ 200 lines have an ACF365 of $\approx 0$.  

These results demonstrate that physically motivated line selection based solely on depth substantially improves RV precision over Random, but additional information beyond depth may be required to identify the most stable lines.

\section{Filtering Lines for Accurate RV Exoplanet Surveys (FLARES) \label{Filtering Lines for Accurate RV Exoplanet Surveys (FLARES)}}

Section \ref{1D Removal based on Spectral Line Depth} showed that iteratively removing the shallowest lines via FLARES$_{\textrm{depth}}$ could reduce the RV RMS and mitigate the effects of stellar activity. 
However, FLARES$_{\textrm{depth}}$ considered only a single line property (depth). 
Other line characteristics such as SNR, detector position, activity sensitivity, and susceptibility to telluric contamination can also influence RV precision.
It is therefore useful to consider removing lines based on multiple metrics simultaneously to further mitigate stellar activity and also mitigate RV variability from the instrument and from telluric contamination. 
This motivates an iterative line selection framework that can rank and remove lines using arbitrary combinations of line properties. 
In this section, we introduce the Filtering Lines for Accurate RV Exoplanet Surveys (FLARES) algorithm. 
We first define the set of metrics used in this work. 

\subsection{Metrics \label{Metrics}}
\begin{figure*}[hbt!]
    \centering
	\includegraphics[width= 18 cm]{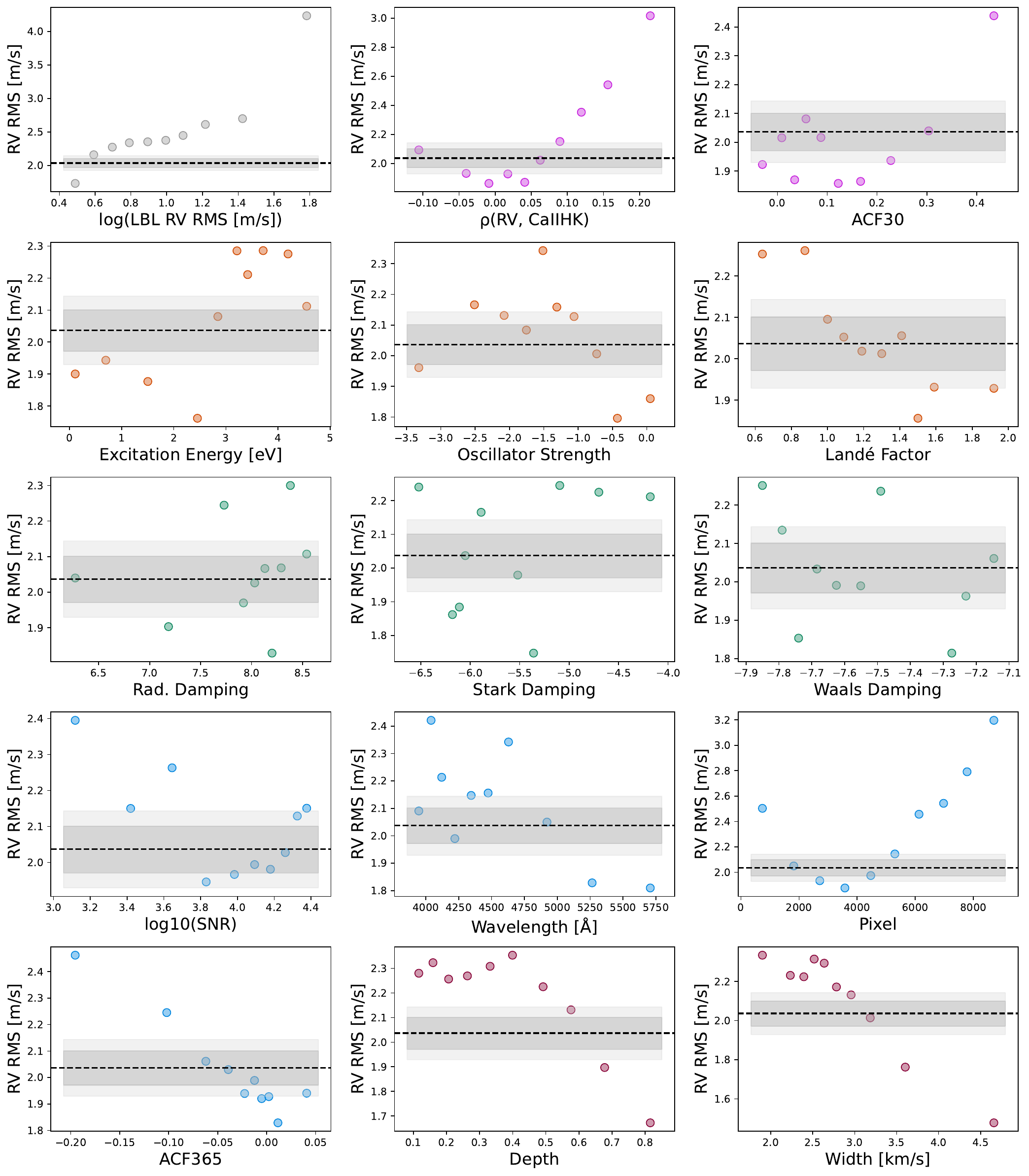}
    \caption{The dependence of RV RMS on various spectral line metrics described in Section \ref{Metrics}. 
    For each metric, all $2,809$ lines are divided into 10 equal bins, such that each bin contains approximately $281$ lines. 
    For comparison, each panel includes the $1\sigma$ (dark gray) and $2\sigma$ (light gray) confidence intervals derived from 1,000 randomly generated line lists with the same size as each bin to illustrate the expected behavior of typical random clusters. 
    The rows approximately describe the metric categories of stellar activity (pink), atomic physics (orange), line broadening parameters (green), instrumental effects (blue), and morphology (red). 
    Many metrics, such as pixel, depth, and $\rho$(RV, Ca\,\textsc{ii}\,HK) show strong trends with RV RMS. 
    Metrics that show weaker trends may still become important in later iterations of FLARES once lines that cause stronger trends have been removed.
    }
    \label{fig:RMS_all_metrics_grid_colored_10bins}
\end{figure*}

In this section, we define metrics by different categories based on whether they trace stellar activity, atomic properties, damping properties, instrumental effects, or line morphology. 
Table \ref{tab:metrics_description} describes the 14 metrics used in this work, and Figure \ref{fig:RMS_all_metrics_grid_colored_10bins} shows the dependence of RV RMS on these metrics using the same binning procedure described for line depth in Section \ref{1D Binning based on Spectral Line Depth}. 
The metrics are colored according to their category.
The LBL RV RMS is also shown in Figure \ref{fig:RMS_all_metrics_grid_colored_10bins} but is \emph{not} used in the FLARES algorithm. 
Figure \ref{fig:RMS_all_metrics_grid_colored_10bins} shows that the bins containing lines with the lowest LBL RV RMS have the lowest overall RV RMS, as expected based on the results from BLARES (see Section \ref{BLARES}). 
We release a dataset of the metric values for each of the 2,809 lines from the curated line list on Zenodo \citep{zenodo_pandey_2026}.

\begin{table*}[hbt]
\caption{Summary of the spectral line metrics used in the FLARES algorithm.}
\label{tab:metrics_description}

\begin{tabular}{||c c c c||}
\hline
\makecell[tl]{Metric} &
\makecell[tl]{Category} &
\makecell[tl]{Unit} &
\makecell[tl]{Data Source} \\[0.5ex]
\hline\hline

\makecell[tl]{$\rho$(RV, Ca\,\textsc{ii}\,HK)} &
\makecell[tl]{Activity} &
\makecell[tl]{N/A} &
\makecell[tl]{LBL RVs and NEID DRP} \\[1ex]
\hline

\makecell[tl]{ACF30} &
\makecell[tl]{Activity} &
\makecell[tl]{N/A} &
\makecell[tl]{LBL RVs} \\[1ex]
\hline

\makecell[tl]{Excitation Energy} &
\makecell[tl]{Atomic} &
\makecell[tl]{eV} &
\makecell[tl]{VALD} \\[1ex]
\hline

\makecell[tl]{Oscillator Strength} &
\makecell[tl]{Atomic} &
\makecell[tl]{N/A} &
\makecell[tl]{VALD} \\[1ex]
\hline

\makecell[tl]{Land\'e factor} &
\makecell[tl]{Atomic} &
\makecell[tl]{N/A} &
\makecell[tl]{VALD} \\[1ex]
\hline

\makecell[tl]{Rad. Damping} &
\makecell[tl]{Damping} &
\makecell[tl]{N/A} &
\makecell[tl]{VALD} \\[1ex]
\hline

\makecell[tl]{Stark Damping} &
\makecell[tl]{Damping} &
\makecell[tl]{N/A} &
\makecell[tl]{VALD} \\[1ex]
\hline

\makecell[tl]{Waals Damping} &
\makecell[tl]{Damping} &
\makecell[tl]{N/A} &
\makecell[tl]{VALD} \\[1ex]
\hline

\makecell[tl]{SNR} &
\makecell[tl]{Instrumental} &
\makecell[tl]{N/A} &
\makecell[tl]{NEID Observations} \\[1ex]
\hline

\makecell[tl]{Wavelength} &
\makecell[tl]{Instrumental} &
\makecell[tl]{\AA} &
\makecell[tl]{VALD} \\[1ex]
\hline

\makecell[tl]{Pixel} &
\makecell[tl]{Instrumental} &
\makecell[tl]{N/A} &
\makecell[tl]{NEID Detector} \\[1ex]
\hline

\makecell[tl]{ACF365} &
\makecell[tl]{Instrumental} &
\makecell[tl]{N/A} &
\makecell[tl]{LBL RVs} \\[1ex]
\hline

\makecell[tl]{Depth} &
\makecell[tl]{Morphology} &
\makecell[tl]{N/A} &
\makecell[tl]{Line list} \\[1ex]
\hline

\makecell[tl]{Width} &
\makecell[tl]{Morphology} &
\makecell[tl]{$\mathrm{km\,s^{-1}}$} &
\makecell[tl]{Line Fitting} \\[1ex]
\hline

\end{tabular}
\end{table*}

\subsubsection{Stellar Activity \label{Stellar Activity}}

\textbf{Correlation with Ca\,II\,HK:} We include the correlation between the LBL RVs and Ca\,\textsc{ii}\,HK, $\rho$(RV, Ca\,\textsc{ii}\,HK). 
For a given day, we use the median Ca\,\textsc{ii}\,HK activity index values from the NEID DRP\footnote{\url{https://neid.ipac.caltech.edu/docs/NEID-DRP/algorithms.html\#stellar-activity-info}} for each daily observation. 
Figure \ref{fig:RMS_all_metrics_grid_colored_10bins} shows that the bins containing lines with no correlation with Ca\,\textsc{ii}\,HK have the lowest RV RMS, and this scatter increases drastically when looking at bins with greater correlations or slight negative correlations with with Ca\,\textsc{ii}\,HK. 

\textbf{ACF30}: We include ACF30 of the LBL RVs, as periodicities occurring at timescales of the solar rotation period could be caused by stellar activity and sunspots/faculae/plage that rotate in and out of view from the Sun's surface. 
Figure \ref{fig:RMS_all_metrics_grid_colored_10bins} shows that the bins containing lines with low ACF30 tend to have lower RV RMS, and the RV RMS increases for bins containing lines with larger ACF30. 

\subsubsection{Atomic Parameters \label{Atomic Parameters}}

All atomic parameters described in this section were obtained from VALD \citep{ryabchikova_2015}.

\textbf{Excitation Energy}: The excitation energy is the energy of the lower level of the transition relative to the ground state.
Lines requiring lower excitation energy tend to be stronger and form higher in the atmosphere, and therefore tend to be deeper lines. 
Figure \ref{fig:RMS_all_metrics_grid_colored_10bins} shows that lines that require less excitation energy show lower RV RMS. 

\textbf{Oscillator Strength}: The oscillator strength gives the transition probability between energy levels and is correlated with the line depth. 
Figure \ref{fig:RMS_all_metrics_grid_colored_10bins} shows that lines with the largest oscillator strength tend to show the lowest RV RMS, though the trend is not as clear as the trend with depth. 

\textbf{Landé Factor}: The Landé $g$-factor measures how strongly a spectral line responds to an external magnetic field from the Zeeman effect. 
Figure \ref{fig:RMS_all_metrics_grid_colored_10bins} shows that lines with the highest Landé $g$-factors, or lines that are more sensitive to magnetic activity, tend to show a lower RV RMS. 
There is an expectation that more magnetically sensitive lines should show a larger RV RMS as they are more sensitive to activity, but it is not obvious why we find the opposite trend here.

\subsubsection{Damping Parameters \label{Damping Parameters}}

All damping parameters described in this section were obtained from VALD \citep{ryabchikova_2015}.

\textbf{Radiative damping}: Radiative damping describes the natural broadening process due to the Heisenberg uncertainty principle.  
Figure \ref{fig:RMS_all_metrics_grid_colored_10bins} shows that there is no clear trend between the radiative damping and the binned RV RMS, though there are some bins that have an RV RMS that deviate significantly from random bootstrapped samples of the same size (e.g., lines with large radiative damping show both large and small RV RMS). 

\textbf{Stark damping}: Stark damping describes broadening due to interactions with charged particles. 
Figure \ref{fig:RMS_all_metrics_grid_colored_10bins} shows that there is no clear trend between the Stark damping and the binned RV RMS. 
%Perhaps there is a slight preference for bins with larger Stark damping to have a larger RV RMS than random bootstrapped samples of the same size. %<- looks like scatter to me.  The  question is why the scatter when dived by Stark is larger than when bins are chosen randomly.  I'm not sure that's worth figuring out. 

\textbf{van der Waals damping}: van der Waals damping describes broadening due to interactions with neutral particles via van der Waals forces. 
Figure \ref{fig:RMS_all_metrics_grid_colored_10bins} shows that there is no clear trend between the van der Waals damping and the binned RV RMS.

\subsubsection{Instrumental Effects \label{Instrumental Effects}}

\textbf{SNR}: For every line and day, we compute the SNR as the ratio of the summed flux to the quadrature sum of the flux uncertainties across the line. 
This quantity is scaled by $\sqrt{N_{\mathrm{exp}}}$, where $N_{\mathrm{exp}}$ is the number of exposures for a given day. 
The final value is taken as the median over all days. 
Figure \ref{fig:RMS_all_metrics_grid_colored_10bins} shows that the RV RMS is minimized when the $3.8 \leq $ log10(SNR) $\leq 4.2$. 
If the RV uncertainties were dominated purely by photon noise, the RV RMS should decrease with increasing SNR. 
However, stellar activity and instrumental effects can also contribute significantly to the RV scatter which can weaken this trend. 
In addition, the RVs are combined using weights derived from the RV uncertainties of each line (see Equation \eqref{EQ:weighted_rv_uncertainty}).
Since lower-SNR lines have larger RV uncertainties, they contribute less to the weighted-average RVs and this makes the relationship between RV RMS and SNR more unclear.

\textbf{Wavelength}: We use the central wavelength of the line from VALD. 
In general, lines in the redder orders ($\lambda \geq 5000$ \AA) have a lower RV RMS, likely because the NEID spectra have higher throughput and SNR in much of the red optical which means they have more available Doppler information \citep{bouchy_2001}.

\textbf{Pixel}: The RV measurements of spectral lines can be systematically affected by their position on the detector, especially near the edges of each physical \'echelle order where instrumental effects such as inaccurate wavelength calibration, illumination variations, or imperfect flat-fielding are more pronounced. 
To quantify this effect, we find the detector pixel that is closest to the central wavelength of each line.
Figure \ref{fig:RMS_all_metrics_grid_colored_10bins} shows that lines that lie near the centers of the \'echelle orders generally exhibit a lower RV RMS than those near the order edges.

\textbf{ACF365}: We include ACF365 of the LBL RVs, as periodicities occurring at timescales of a year could be caused by instrumental effects or telluric contamination due to the seasonal change of airmass. 
Figure \ref{fig:RMS_all_metrics_grid_colored_10bins} shows that the bins containing lines with ACF365 $\approx$ 0 have the lowest RV RMS, and the RV scatter increases as the ACF365 deviates further from 0, especially for slightly negative ACF365. 

\subsubsection{Line Morphology \label{Line Morphology}}

\textbf{Depth}: Sections \ref{1D Binning based on Spectral Line Depth} and \ref{1D Removal based on Spectral Line Depth} already highlighted the importance of line depth on the RV RMS. 
We use depths from VALD \citep{ryabchikova_2015}. 

\textbf{Width}: We measure line widths from Gaussian fits (see Equation \eqref{eq:gaussian_model} in Appendix \ref{Gaussian Model for Line Fitting}). 
The final width value is taken as the median over all days. 
Figure \ref{fig:RMS_all_metrics_grid_colored_10bins} shows that broader lines have a lower RV RMS as they also tend to be deeper. 

\subsection{Iterative Line Removal \label{Iterative Line Removal}}

The next step in the FLARES algorithm is to iteratively remove lines from the full line list.
We initially use all the $N_{\text{metric}} = 14$ metrics for all lines as defined in Section \ref{Metrics}. 
First, we start with all 2,809 lines. 
At each iteration, we select a pool of potential candidate lines to remove for each metric. 
Initially,  $N_{\mathrm{pool}}=10$.
Once there are fewer than 100 lines remaining,  $N_{\mathrm{pool}}$ is set to the ceiling of 10$\%$ of the remaining active lines to avoid having large pool sizes for only few lines.
For each metric, we identify the $N_{\mathrm{pool}}$ lines with largest values for that metric and $N_{\mathrm{pool}}$ lines with smallest values for that metric.
Then, we randomly select one line to become a candidate for removal from each of the minimum and maximum pools for each metric, resulting in $2 N_{\text{metric}}$ candidate lines.
For each candidate line, we temporarily remove it from the current set of active lines and evaluate the RV RMS on the remaining lines using Equation \eqref{EQ:weighted_rv_timeseries}. 
Then, we choose to remove the candidate line that produces the lowest RV RMS after its removal.
This procedure is repeated until only a single line remains. 
%, where $l_m^{(k)}$ is the line with the $k$th smallest value of metric $m$. 

This algorithm is similar to FLARES$_\textrm{depth}$ implemented in Section \ref{1D Removal based on Spectral Line Depth}, but FLARES uses many metrics instead of just depth. 
In addition, instead of evaluating only the single minimum and maximum lines for each metric, we define a pool of extreme candidate lines. 
This is because a metric can become ``stuck'' on a particular extreme-valued line whose removal produces little to no improvement in the RV RMS. 
By randomly selecting candidates from a broader pool of extreme lines, the algorithm is able to explore nearby alternatives and avoid repeatedly testing lines that are not beneficial to remove.

Occasionally, we find that removing any of the candidate lines would \emph{increase} the RV RMS. 
In these cases, we still remove the candidate line that results in the lowest RV RMS after removal.
Future iterations of FLARES may consider redrawing new candidate lines from the pool if a candidate line increases the RV RMS. 

A key limitation of this approach is that it is inherently greedy. 
At each iteration, the FLARES algorithm removes the candidate line that achieves the lowest \emph{immediate} RV RMS, but it is possible that a more optimal subset of lines requires retaining a line that appears to be unfavorable for a particular iteration so that its RV variability is partially canceled by another line. 
For example, FLARES may opt to remove a line with a strong correlation with Ca\,\textsc{ii}\,HK, but its RV signal may be offset by another line with the opposite signal had it been kept. 
Accepting this limitation leads to a dramatic reduction in computational cost relative to a global search. 

Another limitation of this approach is that results from each iteration of FLARES are inherently random. 
It is, however, expected that the highest-performing lines are retained in the later iterations regardless of the random choices made during a given run. 
We explore this further in Section \ref{Monte Carlo Simulations of FLARES} by running 100 Monte Carlo simulations of FLARES to evaluate its robustness. 
We release a dataset of LBL rankings based on when lines were removed by FLARES on Zenodo \citep{zenodo_pandey_2026}. 

\section{Results \label{Results}}

\subsection{FLARES vs Benchmark Methods \label{FLARES vs Benchmark Methods}}
\begin{figure*}[hbt!]
    \centering
	\includegraphics[width= 18 cm]{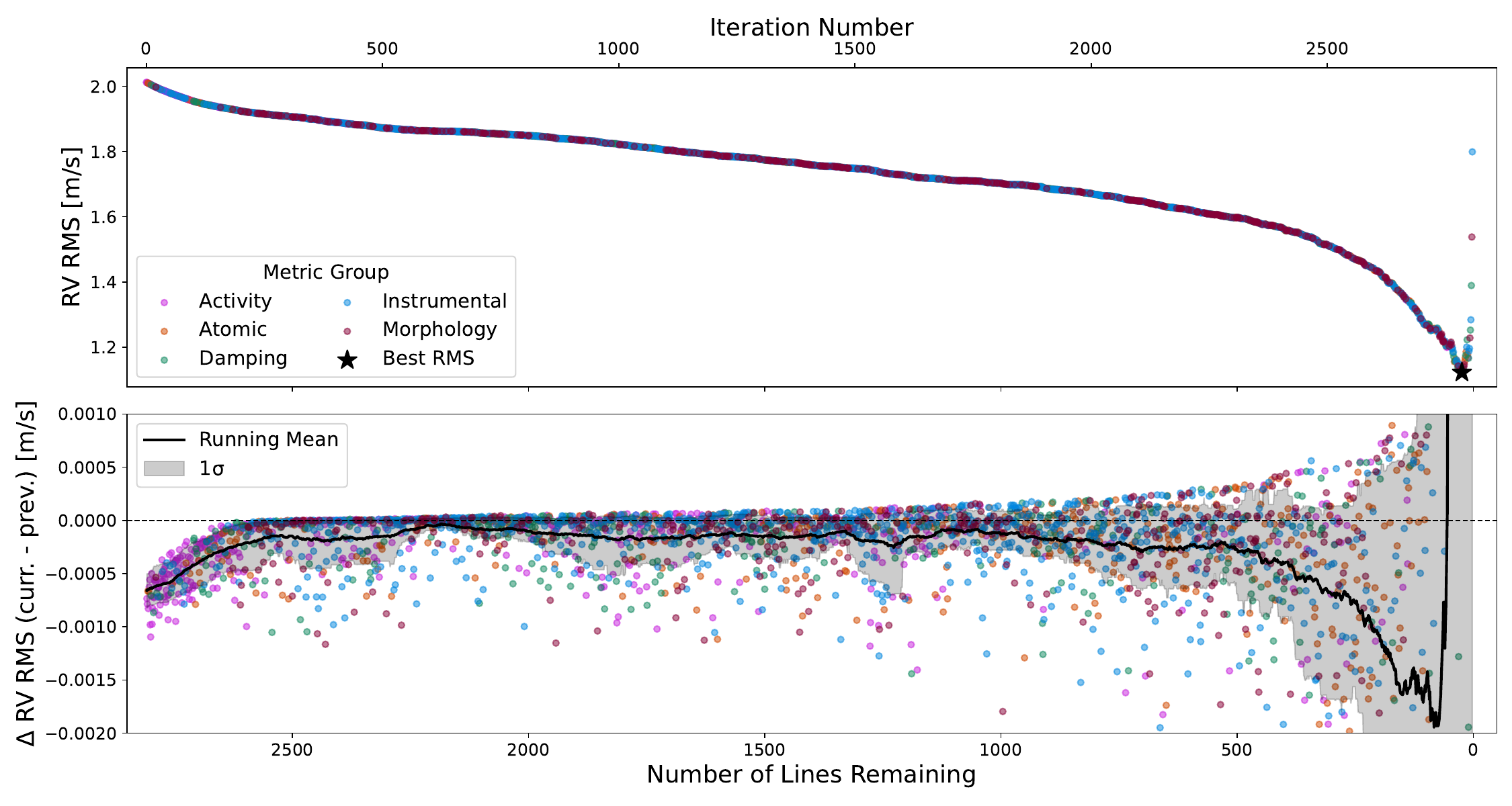}
    \caption{Evolution of the RV RMS as the worst lines are iteratively removed, based on a single run of the FLARES algorithm described in Section \ref{Filtering Lines for Accurate RV Exoplanet Surveys (FLARES)}. 
    \textit{Top:} the RV RMS as a function of the number of lines remaining. 
    Points are colored by the metric category used for the corresponding removal step. 
    The metric categories are the same as those defined in Section \ref{Metrics} and are activity (pink), atomic (orange), damping (green), instrumental (blue), morphology (red).
    The black star marks the minimum RMS of $\approx 1.122~\mathrm{m\,s^{-1}}$ achieved by the algorithm using the best 24 lines. 
    \textit{Bottom:} the change in RV RMS between successive steps, $\Delta \mathrm{RMS} = \mathrm{RMS}_{i} - \mathrm{RMS}_{i-1}$, as a function of the number of lines remaining. 
    Individual steps are colored by the metric category used for removal. 
    The black curve shows a running mean using a 100 iteration window, and the shade depicts the $1\sigma$ dispersion and shows the typical impact of removing a line at different stages of FLARES. 
    Negative values correspond to improvements in RV precision.
    For the first $\approx$ 200 iterations, FLARES chooses lines based on activity metrics such as $\rho$(RV, Ca\,\textsc{ii}\,HK) and ACF30.
    }
    \label{fig:rv_rms_vs_num_lines}
\end{figure*}

In this section, we compare the results of a single run of FLARES with results from the benchmark algorithms described in Section \ref{Benchmark Line Selection Algorithms}.  
Figure \ref{fig:rv_rms_vs_num_lines} shows the result of a single run of the FLARES algorithm using all $N_{\text{metric}} = 14$ metrics described in Section \ref{Metrics}. 
We refer to this singular run as FLARES for the rest of this work, and explore the more general robustness and reproducibility of the FLARES algorithm across 100 Monte Carlo simulations in Section \ref{Monte Carlo Simulations of FLARES}. 

The best 24 lines chosen by FLARES have the lowest RV RMS of $\approx 1.122~\mathrm{m\,s^{-1}}$, as compared to an RV RMS of $\approx 2.012~\mathrm{m\,s^{-1}}$ using all lines. 
Figure \ref{fig:rv_rms_all_methods} shows that FLARES consistently has an RV RMS that is $\sim 0.3~\mathrm{m\,s^{-1}}$ lower than BLARES, suggesting that iteratively removing lines guided by metrics and line properties is a more effective strategy than simply using the LBL RV RMS.  
However, it is still much larger than the photon-noise limit, even for the best 20$-$30 lines. 
This suggests that the metrics used by FLARES do not fully capture all sources of RV variability and that additional information beyond the metrics considered here may be required to approach the photon-noise limit.

\subsection{Metric Contributions to FLARES Performance \label{Metric Contributions to FLARES Performance}}

It is useful to see whether certain metrics or metric categories dominate at particular stages of FLARES. 
Table \ref{tab:flares_metric_weights} shows how often each metric was selected, and the associated weight for each metric when finding nearest-neighbors of each line (see Section \ref{Constructing Comparison Line Lists}). 

\begin{table}[hbt]
\centering
\caption{The number of times each metric was selected for removal by FLARES, $N_m$, along with the corresponding weights, $w_m$ used in the nearest-neighbor analysis for constructing comparison line lists (see Section \ref{Constructing Comparison Line Lists}). 
Weights are defined as $w_m \propto \sqrt{N_m}$ and normalized to unity.}
\label{tab:flares_metric_weights}

\begin{tabular}{l c c}
\hline
Metric & $N_m$ & $w_m$ \\
\hline

Pixel            & 427 & 0.1065 \\
$\rho$(RV, Ca\,\textsc{ii}\,HK) & 360 & 0.0978 \\
SNR                       & 277 & 0.0858 \\
Width                     & 247 & 0.0810 \\
Depth                     & 218 & 0.0761 \\
Stark Damping             & 188 & 0.0707 \\
Land\'e factor            & 175 & 0.0682 \\
Excitation Energy         & 166 & 0.0664 \\
Waals Damping             & 159 & 0.0650 \\
ACF30                     & 155 & 0.0642 \\
Oscillator Strength       & 132 & 0.0592 \\
Rad. Damping              & 111 & 0.0543 \\
ACF365                    & 105 & 0.0528 \\
Wavelength                & 88  & 0.0483 \\
\hline

\end{tabular}
\end{table}

\begin{figure}[hbt!]
    \centering
	\includegraphics[width= 8 cm]{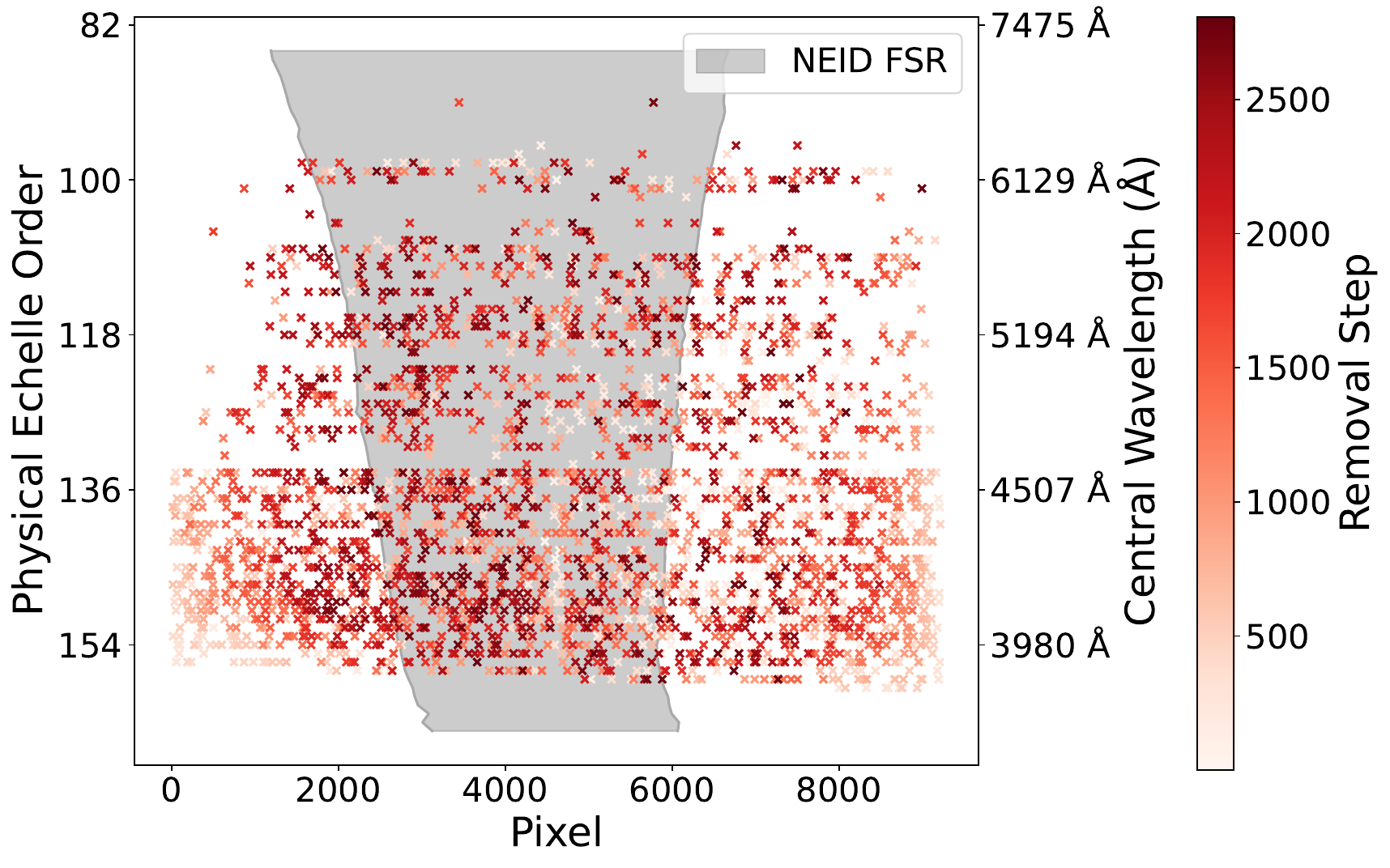}
    \caption{Locations of all lines on the detector, colored by the step they were removed by FLARES. 
    Lines colored a darker red are removed later in FLARES, suggesting these lines yield RVs less affected by stellar and instrumental variability. 
    The gray shade shows an approximate projection of the NEID FSR onto the extracted physical \'echelle orders.
    Lines near the edge of the orders are removed in early iterations of FLARES.
    }
    \label{fig:Detector_greedy}
\end{figure}

Detector location has the strongest influence on FLARES, with the pixel metric selected 427 times during the iterative removal process (see Table \ref{tab:flares_metric_weights}).
Figure \ref{fig:Detector_greedy} shows the locations of all lines on the detector, colored by the step they were removed by FLARES. 
Lines near the edge of the physical \'echelle orders tend to be removed early in the process due to instrumental effects such as inaccurate wavelength calibration, variations in continuum normalization, and imperfect flat-fielding. 
%There is also a slight preference for FLARES to choose to keep lines in bluer orders as compared to the redder orders. 

\begin{figure}[hbt!]
    \centering
	\includegraphics[width= 8 cm]{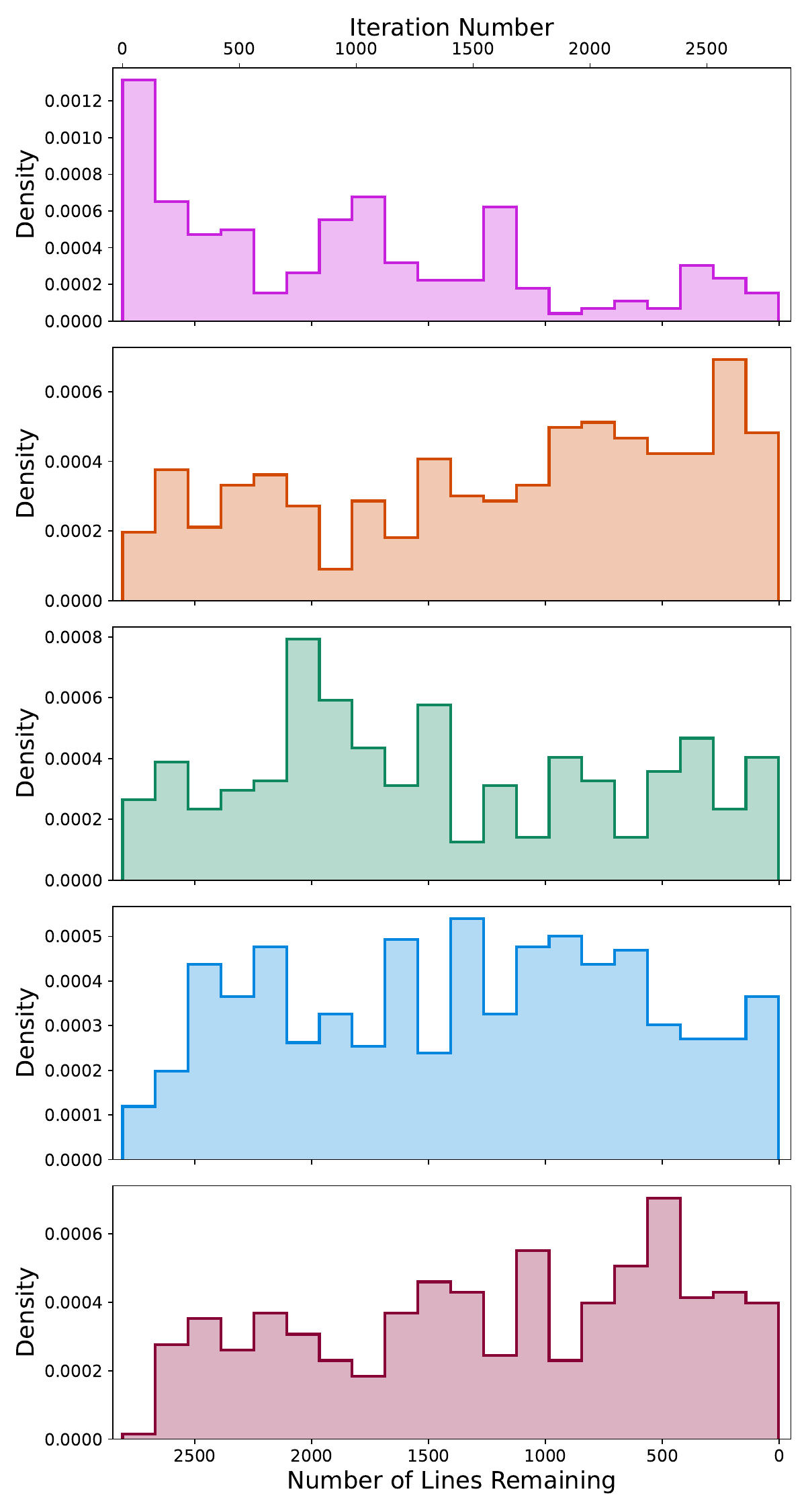}
    \caption{The distribution of the number of FLARES removal events as a function of the number of lines remaining for each metric category. From top to bottom, the metric categories shown are activity (pink), atomic (orange), damping (green), instrumental (blue), and morphology (red). Lines that are removed based on the activity category are preferentially removed in earlier iterations of FLARES, whereas other metric categories are relatively uniform throughout all iterations. 
    }
    \label{fig:metric_group_iteration_hist}
\end{figure}

Table \ref{tab:flares_metric_weights} also shows that correlation with Ca\,\textsc{ii}\,HK has the second-strongest influence on FLARES, as $\rho$(RV, Ca\,\textsc{ii}\,HK) was chosen 360 times. 
Figure \ref{fig:metric_group_iteration_hist} shows that for the first $\approx$ 200 iterations, FLARES chooses lines based on activity metrics such as lines that show a large correlation between RVs and Ca\,\textsc{ii}\,HK and large ACF30. 
This emphasizes the importance of removing those lines most susceptible to stellar activity first. 
Figure \ref{fig:rv_rms_vs_num_lines} also shows that removing these activity-sensitive lines decreases the RV RMS the most as compared to the first $\approx2,500$ iterations, by about $\approx 0.0005~\mathrm{m\,s^{-1}}$ per iteration. 
Although this change is small for any individual step, it is notable because it is achieved while removing only a single line at a time from a set of $\approx 2700$ lines, for which the combined RV RMS is already very precise. 
After hundreds of iterations, the small improvements by removing these activity-sensitive lines decreases the RV RMS from $\approx 2.012~\mathrm{m\,s^{-1}}$ to $\approx 1.9~\mathrm{m\,s^{-1}}$. 

Figure \ref{fig:metric_group_iteration_hist} shows that beyond the first $\approx 200$ iterations, there are no metric categories that dominate particular stages of FLARES. 
There is a slight tendency for atomic and morphology categories to occur at later iterations, but the damping and instrumental categories are largely uniform.  

\begin{figure*}[hbt!]
    \centering
	\includegraphics[width= 18 cm]{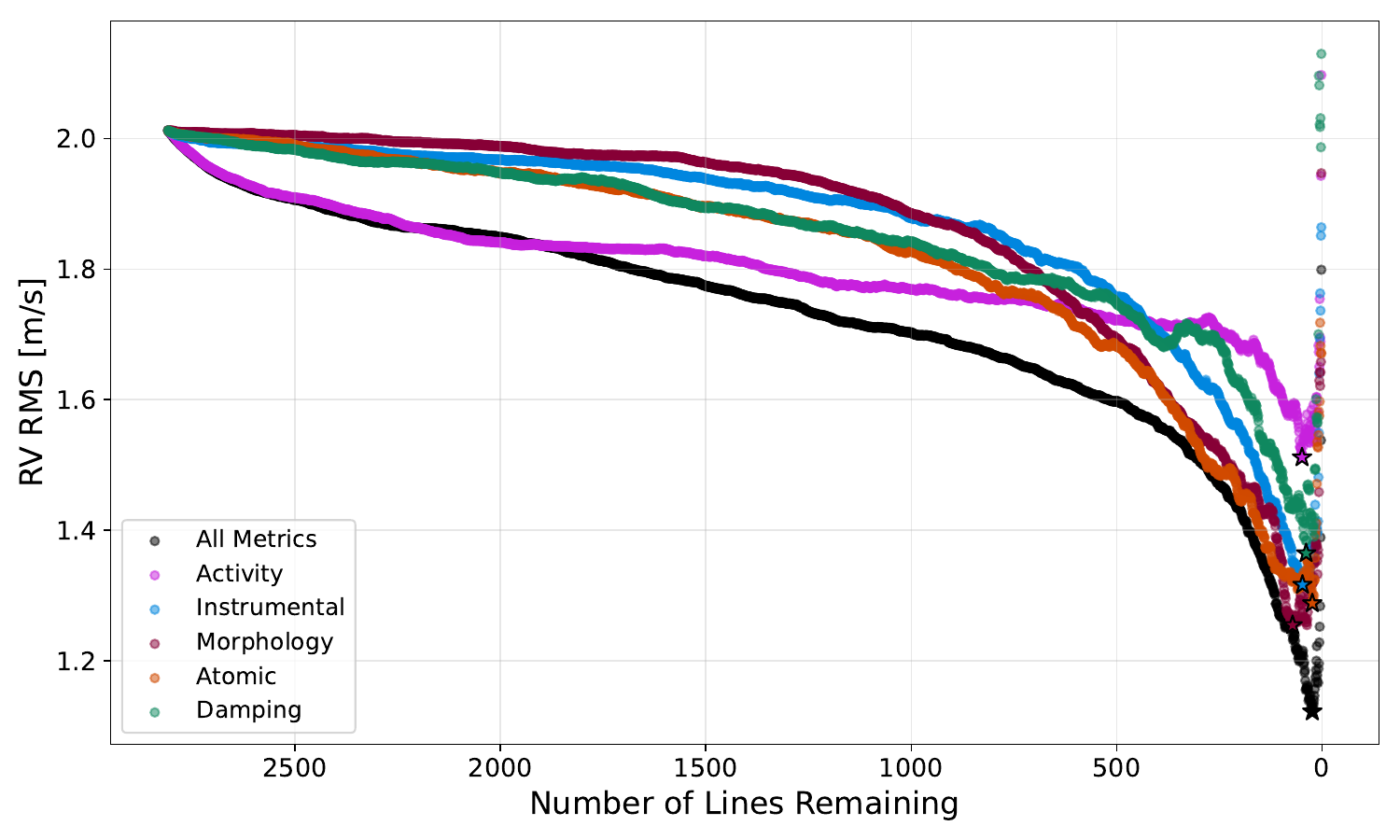}
    \caption{Evolution of the RV RMS as the worst lines are iteratively removed when running FLARES on only the metrics within each metric category (see Section \ref{Metric Contributions to FLARES Performance}). 
    Black points show the evolution of the algorithm using all metrics (same as the top panel of Figure \ref{fig:rv_rms_vs_num_lines}). 
    Also shown are activity (pink), atomic (orange), damping (green), instrumental (blue), morphology (red).
    When FLARES is only applied on activity metrics, the evolution of the RV RMS is similar to the run with FLARES applied on all metrics for the first $\approx 1000$ iterations. 
    However, no single metric category is able to reproduce the FLARES run using all metrics. 
    }
    \label{fig:greedy_metric_group_comparison}
\end{figure*}

To determine whether any of the metric categories alone can achieve a similar RV RMS, we rerun FLARES five times, each time restricting the input metrics to those from a single category among the five categories described in Section \ref{Metrics}, and results for each of these FLARES runs is shown in Figure \ref{fig:greedy_metric_group_comparison}. 
%\textcolor{red}{(Could consider using a Table to list above values).}

When FLARES is only applied on activity metrics, Figure \ref{fig:greedy_metric_group_comparison} shows that the evolution of the RV RMS is similar to the run with FLARES applied on all metrics for the first $\approx 1000$ iterations. 
However, because only two activity metrics are used, the algorithm only reaches a minimum RV RMS of only $\approx 1.5~\mathrm{m,s^{-1}}$.
Applying FLARES using only morphology or atomic metrics perform the best with $\approx 1.25~\mathrm{m\,s^{-1}}$ and $\approx 1.29~\mathrm{m\,s^{-1}}$, respectively, but still do not achieve the $\approx 1.122~\mathrm{m\,s^{-1}}$ from using the algorithm on all metrics. 
Instrumental and damping achieve $\approx 1.32~\mathrm{m\,s^{-1}}$ and $\approx 1.36~\mathrm{m\,s^{-1}}$, respectively. 
These results suggest that no single metric category contains all the information needed to reach the performance of FLARES using all metrics, although it is possible that a combination of the activity category and a secondary category could achieve similar results as using all metrics. 
This is left for future work. 

\subsection{Properties of Best FLARES-selected Lines \label{Properties of Best FLARES-selected Lines}}

\begin{figure*}[hbt!]
    \centering
	\includegraphics[width= 18 cm]{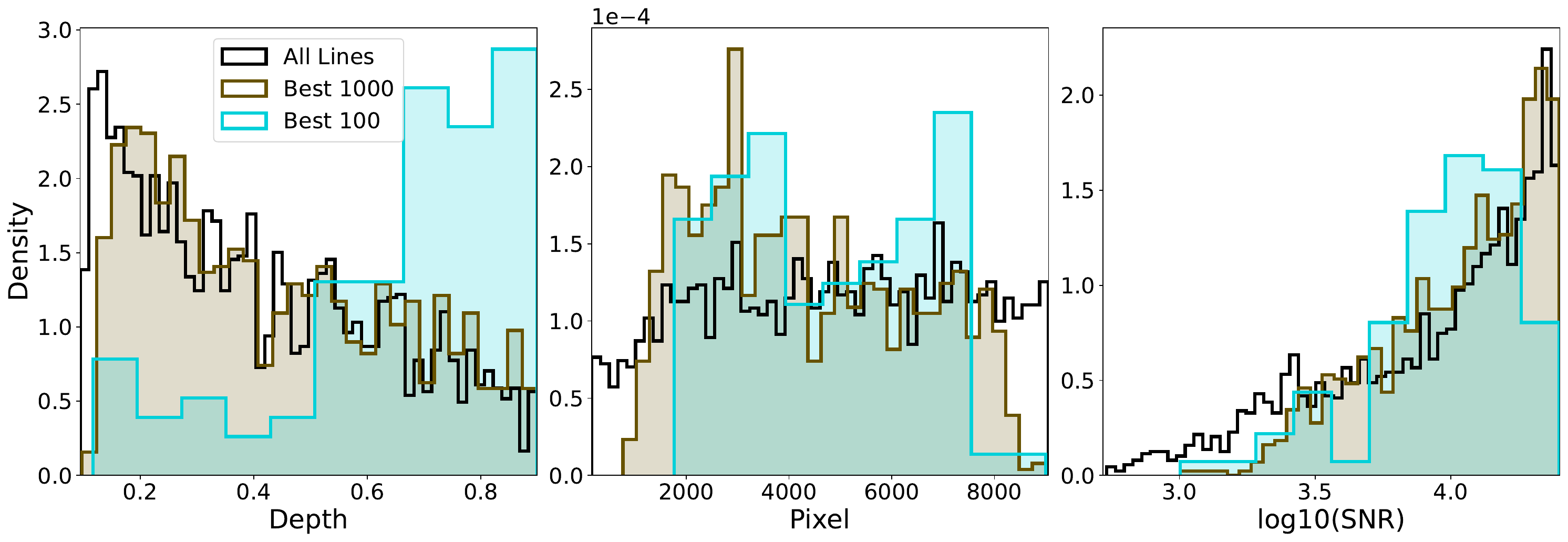}
    \caption{\emph{Left:} the distributions of the depth for all lines (black), for the best 1000 lines (brown), and the best 100 lines (teal) selected by FLARES. 
    \emph{Middle:} same as the left panel, but for the distribution of the pixels of each line. 
    \emph{Right:} same as the left panel, but for the distribution of the SNR of each line. 
    The 100 best lines selected by FLARES tend to be deeper, located near the center of the detector, and have larger SNR as compared to all lines. 
    }
    \label{fig:select_histograms_of_best_lines}
\end{figure*}

It is useful to understand the properties of the best lines chosen by FLARES. 
Figure \ref{fig:select_histograms_of_best_lines} shows that the best 100 lines tend to have a depth $\geq 0.7$, and the best 1000 lines are slightly skewed towards more deeper lines as compared to all lines. 
The best 100 lines are preferentially located near the center of physical \'echelle orders, primarily between pixels 2000 and 7000, whereas the best 1000 lines are distributed somewhat more broadly from pixels 1000 to 7000.
The best 100 lines also tend to have a log10(SNR) $\geq$ 3.7. 
Interestingly, the best 1000 lines seem to actually be skewed towards larger SNR as compared to 100 lines. 
However, both groups of the best 1000 lines and 100 lines are skewed towards higher SNR as compared to the distribution of all lines. 

\begin{comment}
Figure \ref{fig:metric_corner_nlines_100_full_10candidates} in Appendix \ref{Properties of the Best lines from the FLARES Algorithm} shows a corner plot of the distributions of the metrics of all lines, and the distribution of metrics for the 100 best lines chosen by FLARES.
\end{comment}

In addition to the depth, pixel, and SNR discussed above, the best 100 lines exhibit several notable characteristics relative to the full sample:
\begin{enumerate}
\item They show weaker correlations with Ca\,\textsc{ii}\,HK, with a more balanced distribution of positive and negative correlations.
\item They tend to have larger line widths.
\item They exhibit stronger ACF30 signals.
\item They are preferentially located at longer wavelengths, corresponding to redder spectral orders.
\item They show a slight tendency toward more negative ACF365.
\item They exhibit lower individual LBL RV RMS values.
\item Their Land\'e factors show a pronounced peak near 1.5.
\item They have larger oscillator strengths.
\item They generally disfavor the largest excitation energies.
\item They avoid large Stark damping coefficients.
\end{enumerate}

\subsection{Monte Carlo Simulations of FLARES \label{Monte Carlo Simulations of FLARES}}

Since FLARES chooses random candidate lines for each metric at every iteration, different realizations of the algorithm can produce different line rankings and final line lists.
To quantify the robustness and reproducibility of FLARES, we run 100 Monte Carlo (MC) simulations of FLARES, which we refer to as FLARES MC. 

\begin{figure}[hbt!]
    \centering
	\includegraphics[width= 8.5 cm]{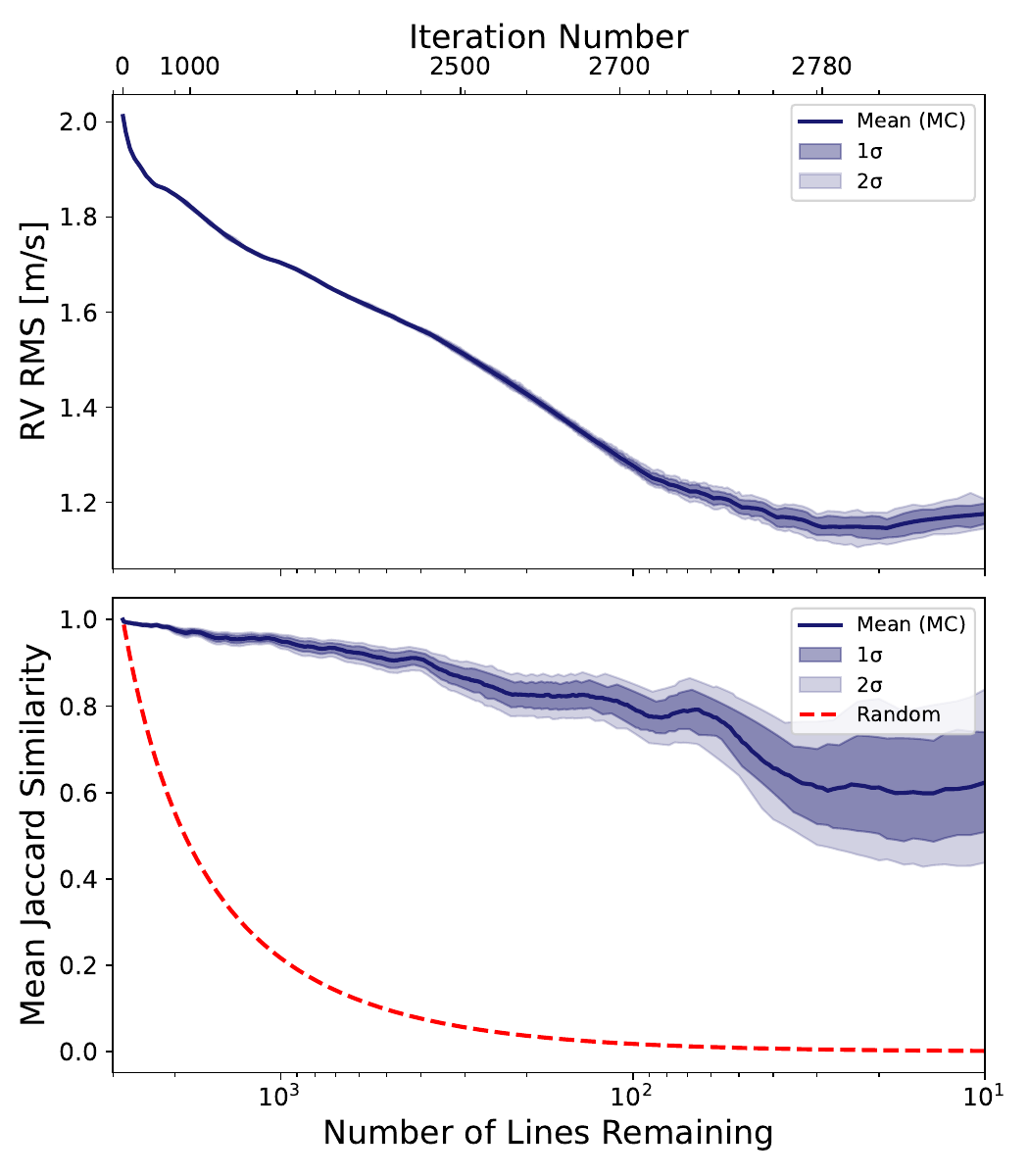}
    \caption{The robustness of FLARES across 100 MC simulations (FLARES MC; see Section \ref{Monte Carlo Simulations of FLARES}). \emph{Top:} the mean RV RMS of FLARES MC as a function of the number of lines remaining. The dark and light blue shaded regions indicate the $1\sigma$ and $2\sigma$ dispersion across FLARES MC, respectively. The small dispersion in RV RMS even with only $20-30$ lines remaining suggests that FLARES is robust across multiple MC simulations. 
    \emph{Bottom:} the mean Jaccard similarity of best line subsets computed across all pairwise MC simulations as a function of the number of lines remaining. The dark and light blue shaded regions indicate the $1\sigma$ and $2\sigma$ dispersion across FLARES MC, respectively. The dashed red curve shows the expected Jaccard similarity for randomly selected subsets of the same size (see Section \ref{Derivation of the Expected Jaccard Similarity}). The mean Jaccard similarity is larger than the expected Jaccard similarity for all sizes suggesting that different MC simulations of FLARES select substantially overlapping (but not identical) subsets. 
    }
\label{fig:FLARES_uncertainty_with_Jaccard_similarity}
\end{figure}

Figure \ref{fig:FLARES_uncertainty_with_Jaccard_similarity} shows the mean RV RMS of FLARES MC along with 1$\sigma$ and 2$\sigma$ dispersion across all MC simulations as a function of the number of lines remaining. 
The 1$\sigma$ dispersion in the RV RMS of FLARES MC is low ($\leq 0.01 ~\mathrm{m\,s^{-1}}$) until there are only $\sim300$ lines remaining. 
The maximum dispersion is reached at the minimum when there are only 20$-$30 lines remaining, and this dispersion is only $\leq 0.1 ~\mathrm{m\,s^{-1}}$. 
This small dispersion indicates that the RV RMS is largely insensitive to the stochastic line selection in FLARES, suggesting that the algorithm is robust and its minimum RV RMS is reproducible.

To quantify the robustness of the line subsets selected by FLARES at each iteration, we find subsets of the best $N$ lines for each MC simulation and compute an overlap between pairs of subsets for each $N$. 
To compute this overlap, we define the Jaccard similarity
\begin{equation}
J(\mathcal{L}_i, \mathcal{L}_j) = \frac{|\mathcal{L}_i \cap \mathcal{L}_j|}{|\mathcal{L}_i \cup \mathcal{L}_j|},
\end{equation}
where $\mathcal{L}_i$ and $\mathcal{L}_j$ are the subsets of the $N$ best lines in simulations $i$ and $j$, respectively, $|\mathcal{L}_i \cap \mathcal{L}_j|$ is the number of lines common to both subsets, and $|\mathcal{L}_i \cup \mathcal{L}_j|$ is the total number of distinct lines contained in either subset. 
A larger $J(\mathcal{L}_i, \mathcal{L}_j)$ occurs when simulations $i$ and $j$ have many common lines. 

We then compute the mean and 1$\sigma$ and 2$\sigma$ dispersion of $J(\mathcal{L}_i, \mathcal{L}_j)$ over all pairs of MC simulations as a function of $N$.
For comparison, we also compute the expected $J(\mathcal{L}_i, \mathcal{L}_j)$ for two independently drawn random subsets of size $N$ from a total of $L=2,809$ lines as $J_{\mathrm{rand}}(N) \approx N/(2L - N)$ (see Appendix \ref{Derivation of the Expected Jaccard Similarity}). 

Figure \ref{fig:FLARES_uncertainty_with_Jaccard_similarity} shows that the mean $J(\mathcal{L}_i, \mathcal{L}_j)$ is close to 1 for large subsets of lines, but decreases to $\approx0.6$ when the minimum RV RMS is typically achieved with about $20-30$ lines. 
This suggests that about 60$\%$ of the lines in these best subsets consistently show up in every MC simulation, whereas the remaining 40$\%$ of lines are random, indicating that different MC simulations of FLARES select substantially overlapping (but not identical) subsets.
Figure \ref{fig:FLARES_uncertainty_with_Jaccard_similarity} also shows that FLARES consistently selects significantly more overlapping subsets as compared to random expectation. 

\subsection{Constructing Comparison Line Lists \label{Constructing Comparison Line Lists}}

Section \ref{Properties of Best FLARES-selected Lines} showed that the best lines selected from FLARES have similar properties, and Section \ref{Monte Carlo Simulations of FLARES} showed that FLARES is robust and reproducable across 100 MC simulations. 
It is therefore natural to ask whether alternative line lists with similar properties can achieve comparable RV precision to FLARES. 
These alternative line lists with matched line properties will provide a valuable comparison to FLARES and the benchmark algorithms from Section \ref{Benchmark Line Selection Algorithms}.  

To construct alternative FLARES line lists, we first find the rank of each line determined by when it is removed in each FLARES MC simulation. 
A lower rank means the line was removed later in FLARES, and is therefore a better-performing line than a line with higher rank. 
We then determine the average and standard deviation of each line's rank across all MC simulations. 
We release a dataset of LBL rankings based on the average rank of when lines were removed via FLARES MC on Zenodo \citep{zenodo_pandey_2026}. 

We then construct a new line subset of the $25$ lines that have the lowest (best) mean rank across FLARES MC, and refer to this as the ``best" subset. 
We choose $25$ lines as $20-30$ lines consistently produced the line subset that minimized the RV RMS for FLARES MC (see Figure \ref{fig:FLARES_uncertainty_with_Jaccard_similarity} and Table \ref{tab:minimum_rv_rms_comparison}). 

For each line in the best subset, we find its nearest neighbors using the following approach. 
First, each spectral line is represented by a vector of all its metric values. 
To compute nearest neighbors, we determine a distance between two lines based on their metrics as the weighted Euclidean distance in metric space. 
When computing the weighted Euclidean distance between lines, we standardize the values of each metric of every line so that the marginal variance is 1. 
We apply weights for each dimension proportional to the square root of the number of times that metric was selected by FLARES for removal (see Table \ref{tab:flares_metric_weights}).
Metric weights were normalized to sum to unity. 

After the weighted Euclidean distances are calculated for every pair of lines, we choose the $k = $ 5 nearest neighbors\footnote{The choice of $k = $ 5 nearest neighbors is arbitrary, but we find that changing $k$ to 3 or 10 nearest neighbors does not significantly alter results at large numbers of lines. For example, when using $k = 10$ nearest neighbors of the best 25 lines, the RV RMS only changes by $\leq 0.05~\mathrm{m\,s^{-1}}$. } for each best line. 
Finally, to construct alternative line lists, we select one line from the set of $k$ nearest neighbors for each of the best lines and compute the RV RMS based on the selected lines. 
We bootstrap 100 such alternative line lists. 
We refer to this as the FLARES MC NN approach. 

\begin{figure*}[hbt!]
    \centering
	\includegraphics[width= 18 cm]{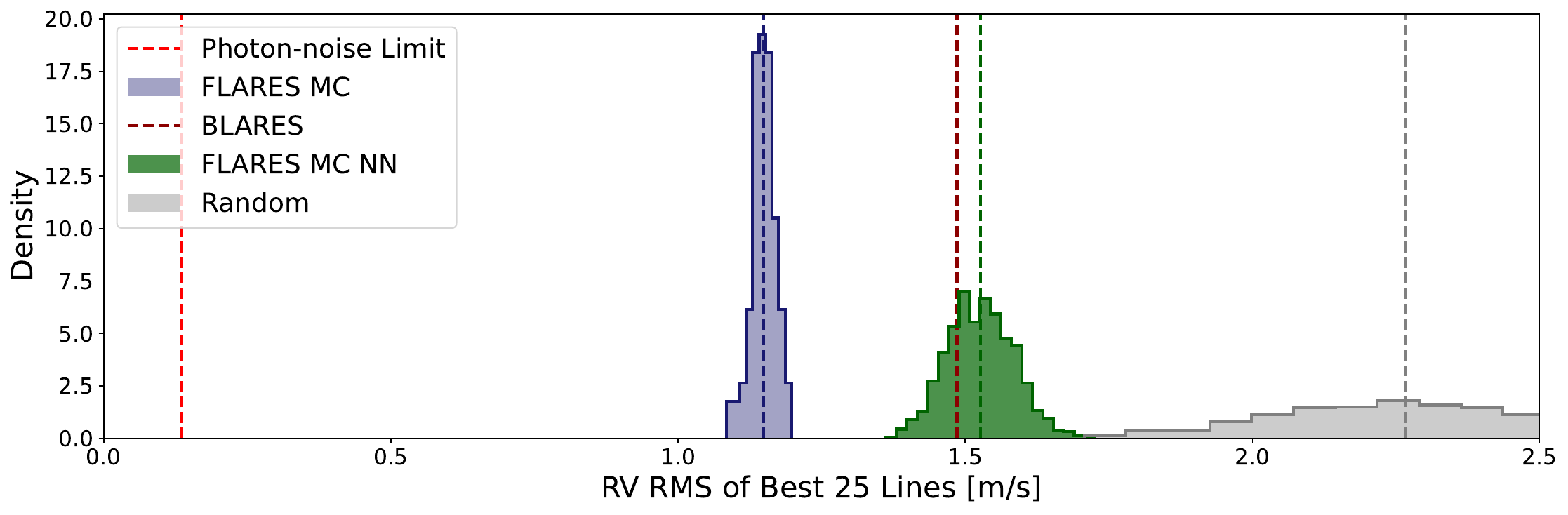}
    \caption{Distributions of the RV RMS of various methods in this work, evaluated for 25 lines. FLARES MC NN (dark-green; see Section \ref{Constructing Comparison Line Lists}) shows the distribution of RV RMS for 100 line lists chosen to have lines that have similar properties to the best lines ranked by FLARES MC. 
    Also shown is the photon-noise limit (dashed red; see Section \ref{Photon-noise Limit}), the distribution of RV RMS for Random (gray; see Section \ref{Random}), the RV RMS of BLARES (dashed dark-red; see Section \ref{BLARES}), and the distribution of RV RMS for FLARES MC (dark-blue; see Section \ref{Filtering Lines for Accurate RV Exoplanet Surveys (FLARES)}), all evaluated for 25 lines. 
    FLARES MC NN performs better than Random and similarly to BLARES, suggesting that constructing alternative line lists based on properties of the 25 best FLARES-selected lines is effective. 
    However, there is a $\sim 0.35~\mathrm{m\,s^{-1}}$ offset between FLARES MC NN and FLARES MC (see discussion in Section \ref{Constructing Comparison Line Lists}). 
    }
    \label{fig:NN_bootstrap_top_25lines_rv_rms}
\end{figure*}

The distribution of the RV RMS of FLARES MC NN is shown in Figure \ref{fig:NN_bootstrap_top_25lines_rv_rms}, along with the photon-noise limit (Section \ref{Photon-noise Limit}), Random (Section \ref{Random}), BLARES (Section \ref{BLARES}), and FLARES MC (Section \ref{Monte Carlo Simulations of FLARES}), all evaluated for 25 lines. 

The distribution of RV RMS for FLARES MC NN is much lower than Random and similar to the RV RMS achieved using BLARES, suggesting that our NN approach is finding subsets of lines with similar properties as FLARES. 
It is expected that FLARES MC NN gives RV RMS values between FLARES MC and Random, as it constructs line lists that try to match the line properties of the best lines from FLARES MC. 
However, Figure \ref{fig:NN_bootstrap_top_25lines_rv_rms} shows that the distribution of RV RMS for FLARES MC NN is still larger than the RV RMS achieved by FLARES MC. 
Ideally, as we add more line metrics and use optimal weights for each metric, we would expect FLARES MC NN to be similar to FLARES MC. 

There could be several reasons for the mismatch between FLARES MC NN and FLARES MC. 
It is possible that due to the inherent ``greedy" nature of FLARES, a portion of the last iterations are choosing lines to remove randomly and not based on any real metric properties. 
This is also suggested by Figure \ref{fig:metric_group_iteration_hist}, where only the activity category preferentially removes lines in the early iterations and the other categories are largely uniform. 
This would suggest that FLARES has already begun overfitting long before only $20-30$ lines are remaining, and in reality FLARES can only achieve $\sim 1.51~\mathrm{m\,s^{-1}}$, which is the minimum achieved using FLARES MC NN using $\approx25$ lines (see Table \ref{tab:minimum_rv_rms_comparison}). 

However, there are still distinct differences in the distribution of line properties between the best 100 and 1000 lines selected by FLARES, as shown by Figure \ref{fig:select_histograms_of_best_lines}. 
This suggests that FLARES is still finding subsets of lines with unique properties even with only $\sim100$ lines remaining. 
This could mean our NN approach is not optimal, either because we have not chosen the best metrics, that we have not weighted the metrics optimally, or we have too many metrics such that it is difficult to find nearest neighbors in a high-dimensional 14-metric space containing only a few thousand lines. 

It is possible that using nearest neighbors of the best subset introduces a bias, since lines in this neighborhood are likely to share similar properties. 
To test this, we repeat the FLARES MC NN procedure while explicitly excluding lines belonging to the best subset from the candidate pool of nearest neighbors. 
This leads to a slight increase in the RV RMS of the resulting alternative line lists, by approximately $\sim 0.1~\mathrm{m,s^{-1}}$.

Finally, Figure \ref{fig:NN_bootstrap_top_25lines_rv_rms} only shows the distributions of RV RMS of the methods used in this work at a snapshot of the best 25 lines. 
Figure \ref{fig:rv_rms_all_methods} shows how the distribution of FLARES MC NN compares with all other methods for various $N$ best lines.
FLARES MC and FLARES MC NN are similar for line lists that are larger than $N\geq1000$ and perform better than BLARES. 
However, it seems that below $N\leq100$, BLARES begins to show a lower RV RMS as compared to FLARES MC NN suggesting that if constructing line lists based on properties of the best lines, it is beneficial to use more than 100 lines.

\section{Conclusions \label{Conclusions}}

In this work, we measured the LBL RVs of 3,830 lines across 383 days of NEID solar observations. 
Using RV information from all lines gave an RV RMS of $\approx 2.012~\mathrm{m\,s^{-1}}$. 
We first tried several benchmark algorithms to select subsets of lines that achieve a lower RV RMS as compared to all lines. 
This included iteratively combining the lines with lowest RV RMS and combining the deepest lines. 
However, neither of these achieved an RV precision close to the limit set by photon-noise. 

Using the benchmark methods as motivation, we introduced the Filtering Lines for Accurate RV Exoplanet Surveys (FLARES) algorithm which iteratively generates line lists while considering multiple line shape metrics and physical properties. 
A single run of FLARES achieved an RV RMS of $\approx 1.122~\mathrm{m\,s^{-1}}$ for NEID Solar RVs using just 24 lines. 
Our results showed that the most significant benefit comes from removing lines most susceptible to stellar activity, such as lines that have RVs most correlated with Ca\,\textsc{ii}\,HK or large periodicities near the solar rotation period. 
Other metrics were largely removed uniformly throughout all the iterations of the algorithm. 
We found that as compared to the full line list, the best 100 lines chosen by FLARES are deeper, have larger SNR, and lie closer to the center of the detector. 

We ran 100 Monte Carlo simulations of FLARES to quantify its robustness and reproducibility. 
We found that across all simulations, FLARES tended to converge to a similar RV RMS and retained the same best lines at the later iterations. 
Finally, we also constructed subsets of line lists that had lines with similar properties to the best lines selected by FLARES using a nearest-neighbor approach.
These alternative line lists had an RV RMS that was better than Random, similar to BLARES, but did not perform as well as line lists generated by FLARES. 

This work raises additional interesting directions for future exploration. 
The FLARES framework provides a general, automated approach to construct line lists that can be applied beyond the Sun. 
Since stellar atmospheres and activity signals vary with spectral type, the optimal set of lines is expected to change from star to star. 
This highlights a limitation of traditional CCF-based approaches which sometimes use fixed line lists and weights that are often transferred across different stars. 
FLARES can provide a more flexible line selection method to be used across different stellar types. 

This work also had several limitations. 
The line selection from FLARES is based on high-cadence and high-SNR daily-averaged solar observations, and it is not yet clear how performance scales to lower-SNR regimes that are typical of nighttime stellar observations. 
Future work could explore how FLARES performance varies as a function of SNR, temporal sampling, and stellar type, and whether the same optimal line lists emerge under more observationally realistic conditions.

FLARES completely discards information from the lines removed in early iterations. 
However, these rejected lines may still contain useful information for diagnosing stellar activity. 
For example, they could potentially be used to correct for activity-induced signals, analogous to previous work showing that the differential behavior between shallow and deep lines can serve as a proxy for stellar activity \citep{cretignier_2020}.
This is a promising direction for future work.

Finally, we also only consider 14 specific metrics in this work. 
Future studies could explore adding additional metrics for selecting candidate lines for removal.  
%Additionally, it would be informative to quantify how well the best lines selected by FLARES based on solar data perform on other G2-type stars and/or stars with more distinct effective temperatures. 

\begin{acknowledgements}
This material is based upon work supported by the National Science Foundation Graduate Research Fellowship Program under Grant No. DGE1255832. 
Any opinions, findings, and conclusions or recommendations expressed in this material are those of the author(s) and do not necessarily reflect the views of the National Science Foundation.

This material is based upon work supported by the U.S. National Science Foundation under grant No. 2204701.

The Center for Exoplanets and Habitable Worlds and the Penn State Extraterrestrial Intelligence Center are supported by Penn State and its Eberly College of Science. 
The authors recognize the Penn State Institute for Computational and Data Sciences (ICDS) (RRID:SCR\_025154) for providing access to computational research infrastructure (RRID:SCR\_026424).
%Computations for this research were performed on the Penn State’s Institute for Computational and Data Sciences' Advanced CyberInfrastructure (ICDS-ACI). 
%This content is solely the responsibility of the authors and does not necessarily represent the views of the Institute for Computational and Data Sciences.

Based on observations at Kitt Peak National Observatory, NSF's NOIRLab, managed by the Association of Universities for Research in Astronomy (AURA) under a cooperative agreement with the National Science Foundation. 
The authors are honored to be permitted to conduct astronomical research on Iolkam Duág (Kitt Peak), a mountain with particular significance to the Tohono Oódham.

Data presented herein were obtained at the WIYN Observatory from telescope time allocated to NN-EXPLORE through the scientific partnership of the National Aeronautics and Space Administration, the National Science Foundation, and the US National Optical-Infrared Astronomy Research Laboratory.
We thank the NEID Queue Observers and WIYN Observing Associates for their skillful execution of our NEID observations. 
Deepest gratitude to Zade Arnold, Joe Davis, Michelle Edwards, John Ehret, Tina Juan, Brian Pisarek, Aaron Rowe, Fred Wortman, the Eastern Area Incident Management Team, and all of the firefighters and air support crew who fought the recent Contreras fire. Against great odds, you saved Kitt Peak National Observatory.

This research has made use of NASA's Astrophysics Data System Bibliographic Services.

This work was performed for the Jet Propulsion Laboratory, California Institute of Technology, sponsored by the United States Government under the Prime Contract 80NM0018D0004 between Caltech and NASA.

\end{acknowledgements}

%\section{Acknowledgments}

\facilities{WIYN (NEID)}

\software{
EchelleCCFs.jl v0.2.9 (\url{https://github.com/RvSpectML/EchelleCCFs.jl}),
EchelleInstruments.jl v0.2.11 (\url{https://github.com/RvSpectML/EchelleInstruments.jl}),
RvLineList.jl (\url{https://github.com/RvSpectML/RvLineList.jl}),
RvSpectML.jl v0.2.8 (\url{https://github.com/RvSpectML/RvSpectML.jl})
}

\appendix \label{appendix}

\section{Comparisons between NEID v1.3 and v1.4 Data \label{Comparisons between NEID v1.3 and v1.4 Data}}

\begin{figure*}[hbt!]
    \centering
	\includegraphics[width=18 cm]{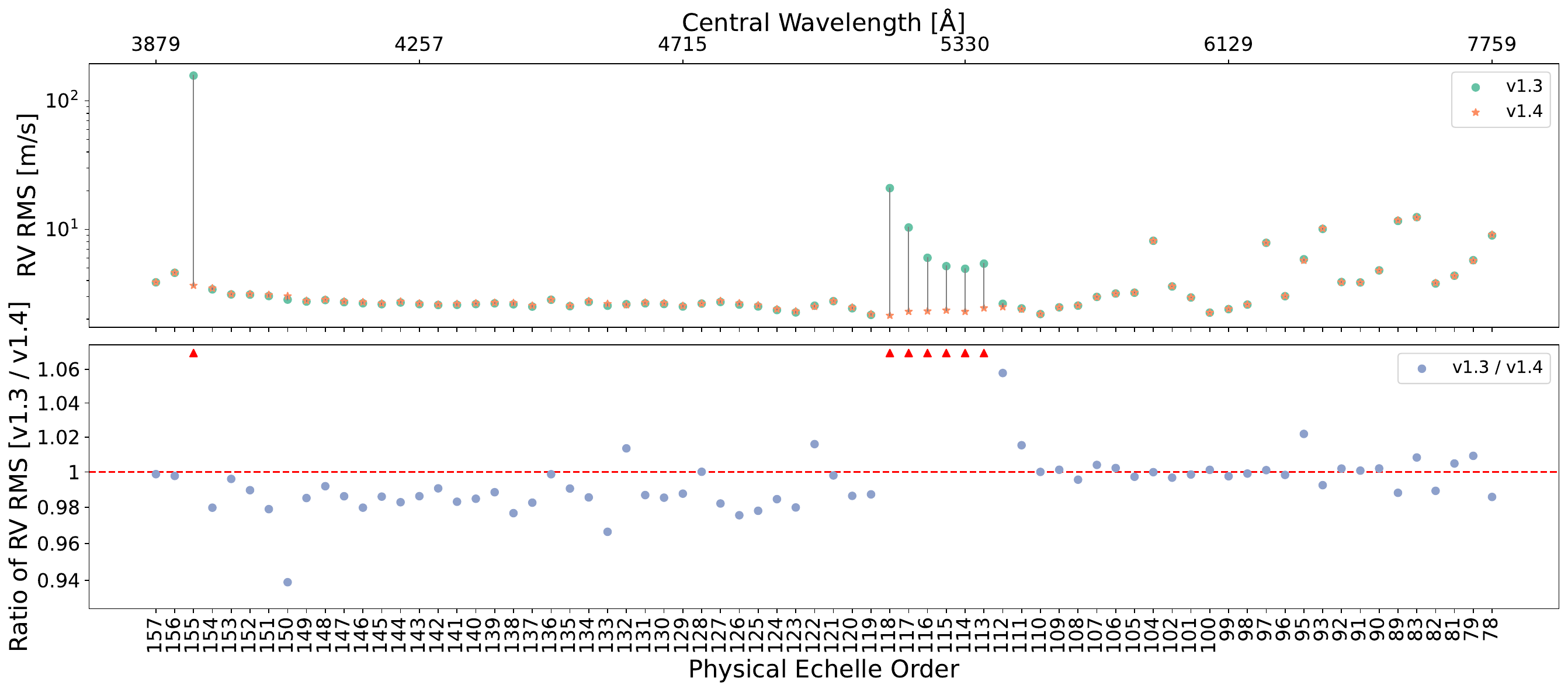}
    \caption{Comparison of the RV RMS across NEID physical \'echelle orders for DRP v1.3 and v1.4. 
    \textit{Top:} the RV RMS for every physical \'echelle order for v1.3 data (teal circles) and v1.4 data (orange stars). 
    The RV RMS was calculated using the PSRP and a Gaussian fit on CCFs across the best days where v1.3 and v1.4 data were both available. 
    Physical \'echelle orders 113$-$118 and 155 have a much lower RV RMS in v1.4 data as compared to v1.3 data. 
    Other physical \'echelle orders show relatively similar RV RMS between v1.3 and v1.4 data. 
    \textit{Bottom:} the ratio between the RV RMS v1.3 and RV RMS v1.4 data for every physical \'echelle order, zoomed in to unity. 
    Red arrows indicate points that are off the limits of the plot. 
    There is a slight tendency for the bluer physical \'echelle orders ($\geq$120) to have a smaller RV RMS from v1.3 data as compared to v1.4 data.
    }
    \label{fig:RV_RMS_by_order_gauss}
\end{figure*}

Since the transition from DRP v1.3 to v1.4 introduced changes to the wavelength solution, we first compare the RV performance of the two reductions to assess the impact of the update. 
Figure~\ref{fig:RV_RMS_by_order_gauss} shows the RV RMS for every physical \'echelle order across all v1.3 and v1.4 observations, calculated using the Penn State Research Pipeline (PSRP) and a
Gaussian fit on CCFs across the best days where v1.3 and v1.4 data were both available. 
DRP v1.4 updated the algorithms used to derive the wavelength calibrations after the LFC was upgraded with a new PCF\footnote{\url{https://neid.ipac.caltech.edu/docs/NEID-DRP/versions.html}}. 
This reduces the RV RMS for physical \'echelle orders 113$-$118 and 155. 
However, there is a slight increase in the RV RMS between v1.3 and v1.4 at the percent level, especially for bluer physical \'echelle orders ($\geq$ 120).

\section{Vetting of NEID Observations} \label{Vetting of NEID Observations}

Initially, there are a total of 258,799 exposures over 1,157 unique days across all three runs. 
We use similar weather filters as in \citet{ford_2024} to choose the best exposures. 
We first filter those exposures which do not pass overall data quality assessments.\footnote{\url{https://neid.ipac.caltech.edu/docs/NEID-DRP/bitfields.html}} 
We also exclude dates with known hardware issues.

\begin{figure*}
    \centering
	\includegraphics[width=18 cm]{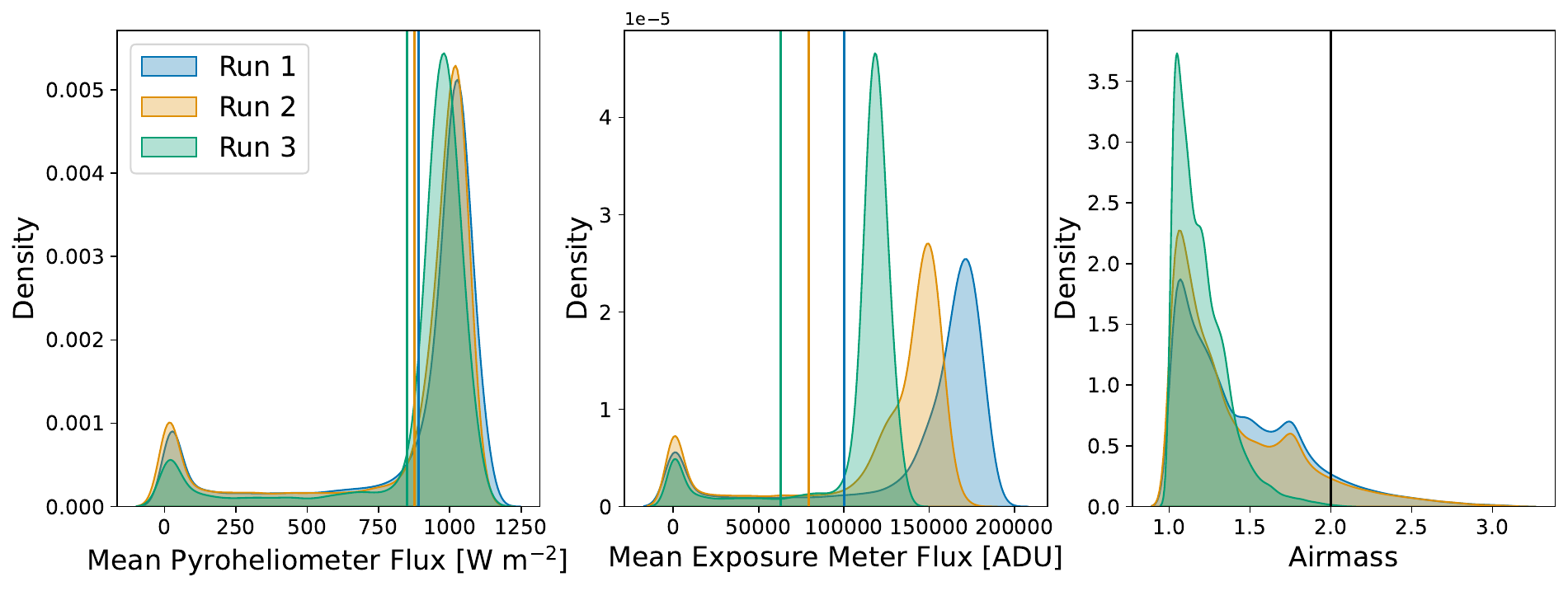}
    \caption{The distribution of the mean pyrheliometer flux (left), mean exposure meter flux (middle), and airmass (right) for all exposures. 
    The distributions are colored blue, yellow, and green for runs 1, 2, and 3, respectively. 
    Vertical colored lines indicate the threshold for each run, and the black line indicates the airmass threshold for the right panel.}
    \label{fig:observation_vetting}
\end{figure*}

We then impose a minimum flux (averaged over the observation) from the pyrheliometer and the exposure meter. 
Since the conditions of the instrument varied between runs, we select flux thresholds for each run such that they filter approximately the same proportion of exposures among the runs, and these thresholds are listed in Table \ref{tab:observation_vetting}. 
Figure~\ref{fig:observation_vetting} shows the distributions of the mean pyrheliometer and exposure meter flux for all exposures in each run, and the corresponding threshold for each run. 

Similar to \citet{ford_2024}, we also require that the mean exposure meter flux, expmeter$\_$mean, and the RMS deviation from the mean of the exposure meter flux, expmeter$\_$rms, for each exposure satisfy
\begin{equation}
    \text{expmeter$\_$rms} \leq 0.003 \times \text{expmeter$\_$mean}.
\end{equation}
Similarly, we require that the mean pyrheliometer flux, pyrflux$\_$mean, and the RMS deviation from the mean of the pyrheliometer, pyrflux$\_$rms, for each exposure satisfy
\begin{equation}
    \text{pyrflux$\_$rms} \leq 0.0035 \times \text{pyrflux$\_$mean}.
\end{equation}
This ensures that flux values do not significantly fluctuate throughout exposures within a single day and removes exposures affected by variable observing conditions that can introduce spurious RV signals. 

We also require that expmeter$\_$mean/pyrflux$\_$mean exceed 95$\%$ of the median value for that ratio within each day to check for pointing errors, and that exposures have an airmass less than 2.0. 
Including exposures taken at an airmass greater than 2.0 increases the effect of differential extinction on the resolved disk of the Sun. 
Lowering the airmass cut below 2.0 would significantly reduce the number of exposures. 

Of the remaining exposures, we finally filter all exposures that occur on days with less than 180 total exposures. 
This threshold allows us to focus on days with the most exposures based on the above criteria. 
This gives us a total of 87,842 exposures over a total of a 383 days.

We adopt the standard NEID wavelength solution provided by the NEID DRP and do not perform any additional drift correction. The spectra are analyzed using the pipeline-provided wavelength calibration which accounts for the daily RV drift of the instrument, as explained in Section 2.1.1 in \citet{ford_2024} and outlined in the NEID DRP manual.\footnote{\url{https://neid.ipac.caltech.edu/docs/NEID-DRP/algorithms.html\#daily-wavelength-calibration}}

\section{Line List Construction \label{Line List Construction}}
In this work, we use RvSpectML's default line list\footnote{\url{https://github.com/RvSpectML/RvLineList.jl}} (Wise et al. in prep.). 
This line list was retrieved from VALD \citep{ryabchikova_2015} with stellar parameters matching the Sun. 
We set the ``extract stellar" star parameters to: Detection Threshold: 0.001, Microturbulence: 1.0, T$_\textrm{eff}$: 5778 K, log g:
4.44, Chemical composition: Fe: $-4.54$. 
The starting and ending wavelengths span from 3000 to 10000 \AA. 
In practice, we retrieved the following wavelength ranges: $3000-4000$ \AA, $4000-6000$ \AA, $6000-9000$ \AA, $9000-10000$ \AA, and concatenated the files (Wise et al. in prep.). 
This line list was generated using the following selection criteria (Wise et al. in prep.):

\begin{itemize}
    \item \texttt{allowBlends = 0}: retain only isolated lines and reject lines with identified blends.

    \item \texttt{overlapCutoff = $5\times10^{-6}$}: lines separated by less than this threshold (expressed as a fraction of the speed of light) are treated as blended.

    \item \texttt{blend\_RV\_cutoff = 2.0}: require the predicted RV perturbation from nearby VALD lines due to activity-induced line-shape changes to be less than $2~\mathrm{m\,s^{-1}}$.

    \item \texttt{depthCutoff = 0.02}: require the line depth to be greater than 0.02 relative to the continuum.
    
    \item \texttt{rejectTelluricSlope = 2000}: reject spectral regions whose synthetic telluric model from TAPAS (Transmissions of the AtmosPhere for AStromomical data; \citealp{bertaux_2014}) has a slope exceeding this threshold.
    
    \item \texttt{quant = 95}: retain only lines whose Gaussian fits converged for at least 95\% of epochs.
    
    \item \texttt{badLineFilter = none}: no additional empirical bad-line mask was applied.

\end{itemize}

For a more thorough explanation of these parameters, see \url{https://github.com/RvSpectML/RvLineList.jl}. 

\section{Window Type} \label{Window Type}

The selection of the window around each line is important in deriving shape parameters and LBL RVs because a large window can produce spurious and/or redundant shape measurements due to neighboring lines, and a small window will only compute shape parameters for the core of the line which is poorer in Doppler information than a window that extends to the wings \citep{cretignier_2020}. 

In this work, we adopt a fixed window extending $\pm 15$ pixels around the line core. 
For NEID, with a resolving power of $R \approx 120{,}000$ and approximately 5 pixels per resolution element \citep{schwab_2016}, this corresponds to $\sim 500~\mathrm{m\,s^{-1}}$ per pixel, or $\pm 7.5~\mathrm{km\,s^{-1}}$ about the line center. 
Since typical solar absorption lines have Full Width at Half Maxima (FWHM) of approximately $7~\mathrm{km\,s^{-1}}$, this window spans roughly two FWHMs in total. 
Although the pixel window is fixed, the physical span in velocity varies with wavelength and dispersion resolution from order to order.

\begin{figure}[htb!]
    \centering
	\includegraphics[width=8 cm]{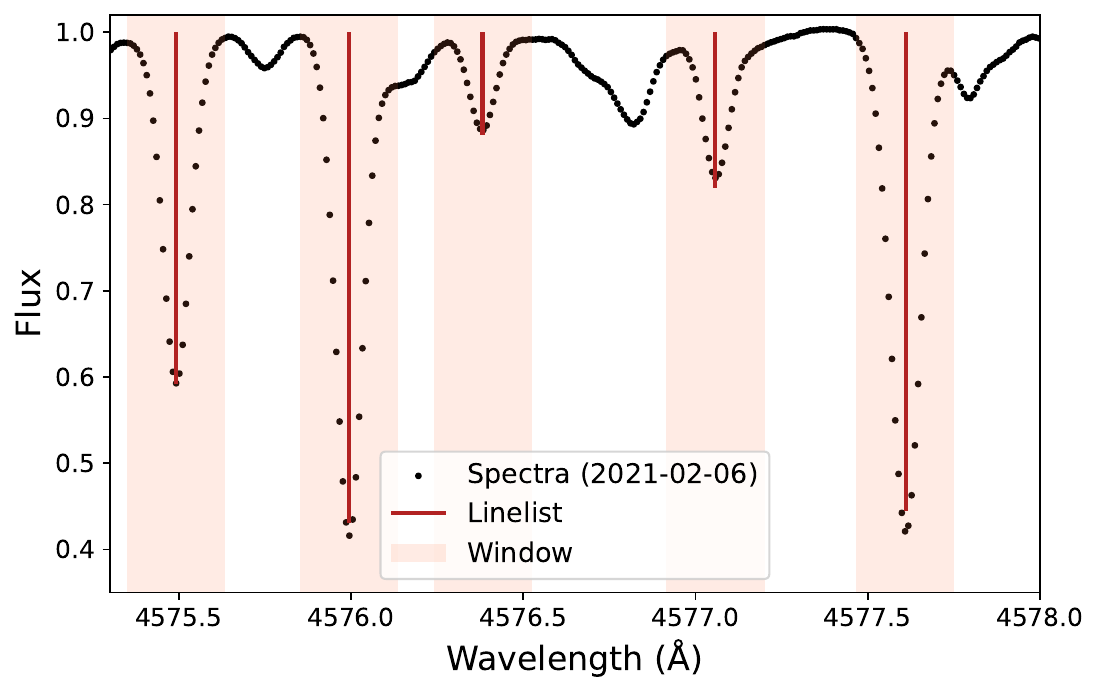}
    \caption{Segment of the daily-averaged NEID solar spectrum from 2021-02-06 (black points) overlaid with spectral lines from the line list (red vertical lines) along with their corresponding 15-pixel window used in this work (orange shade).
    }
    \label{fig:15_pixel_window}
\end{figure}

We note that this window may not be the best window for every line and may be susceptible to blends for narrow lines. 
An alternative approach could be to explore windows that vary for different lines. 
Using alternative windows is left for future work.

\section{Gaussian Model for Line Fitting} \label{Gaussian Model for Line Fitting}

\begin{figure}
    \centering
	\includegraphics[width=8 cm]{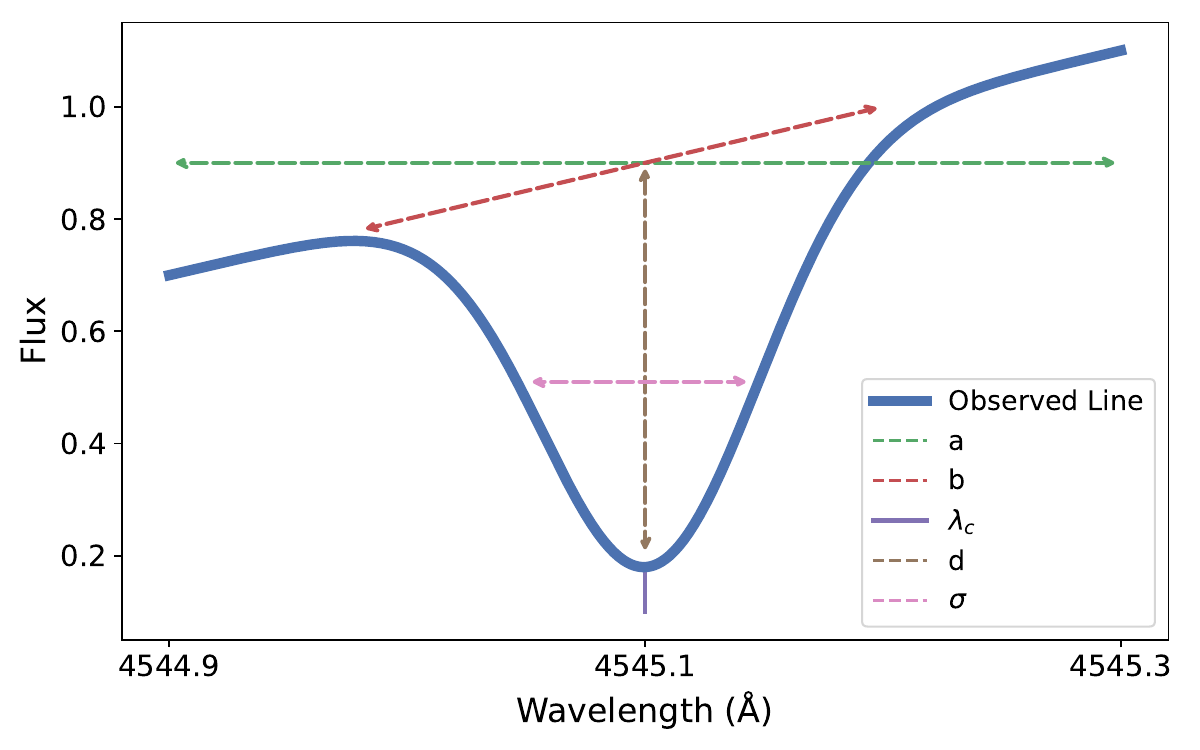}
    \caption{The definitions of line-fitting parameters used throughout this work. 
    A synthetic observed line (solid blue) is shown, and the corresponding Gaussian parameters that describe the observed line are shown as dashed lines and include the continuum level ($a$, green), the continuum slope ($b$, red), the central wavelength ($\lambda_c$, purple), the depth ($d$, brown), and the width ($\sigma$, pink). 
    }
    \label{fig:parameter_definitions}
\end{figure}

For each line, we determine the central wavelength of the line using a Gaussian fit which is used to compute the RV. We use a least squares method to fit a Gaussian model (with a continuum component) to every line:
\begin{equation}
F_{\mathrm{G}}(\lambda)
=
\left[a + b\left(\lambda - \lambda_c\right)\right]
\Bigl[
1 - d\exp\!\Bigl(
-\frac{\bigl[\lambda - \lambda_c(1+z)\bigr]^2}{2\sigma^{2}}
\Bigr)
\Bigr],
\label{eq:gaussian_model}
\end{equation}
\begin{comment}
\begin{equation}
\begin{aligned}
F_{\mathrm{G}}(\lambda)
&= \left[a + b\left(\lambda - \lambda_c\right)\right] \\
&\quad\times
\Bigl[
1 - d\exp\!\Bigl(
-\dfrac{\bigl[\lambda - \lambda_c(1+z)\bigr]^2}{2\sigma^{2}}
\Bigr)
\Bigr],
\end{aligned}
\label{eq:gaussian_model}
\end{equation}
\end{comment}
where $a$ is the continuum level, $b$ is the continuum slope, $\lambda_c$ is the central wavelength, $d$ is the depth, $\sigma$ is the width, and $z$ is the redshift taken as the median redshift given by the NEID DRP L2 file headers for all exposures that day that passed the data quality and weather checks. 
For an initial guess, we use $a = 1$, $b = 10^{-8}$, $\lambda_c$ and $d$ from the line list, and 
\begin{equation}
    \sigma = \frac{v_{\sigma}}{c}\,\lambda_c,
\end{equation}
where $v_{\sigma} = 7$ km~s$^{-1}$ is based on the empirical Full Width at Half Maximum (FWHM) velocity dispersion of solar lines. 
Figure~\ref{fig:parameter_definitions} shows the definitions of these Gaussian parameters on a synthetic line. 

In order to estimate the uncertainties of each shape parameter due to the photon noise for any given line and daily-averaged spectrum, we simulate 100 independent and identically distributed realizations of each line by injecting it with photon noise
\begin{equation}
    F^{\mathrm{noisy}}_i = F^{\mathrm{obs}}_i + \frac{\epsilon_i}{\sqrt{N_{\mathrm{exp}}}}, 
\quad \epsilon_i \sim \mathcal{N}\!\left(0,\,\sqrt{\mathrm{Var}_i}\right)
\end{equation}
where $F^{\mathrm{obs}}_i$ is the observed flux at pixel $i$, $F^{\mathrm{noisy}}_i$ is the simulated, noisy flux at pixel $i$, $\epsilon_i$ is a random variable drawn from a Gaussian distribution with mean 0 and standard deviation $\sqrt{\mathrm{Var}_i}$, where $\mathrm{Var}_i$ is the mean observed variance for pixel $i$ within that day, and $N_{\mathrm{exp}}$ is the number of NEID exposures taken that day that passed the vetting procedure from Section \ref{Vetting of NEID Observations}. 
Uncertainties are then estimated by half the difference between the 84$^\text{th}$ and 16$^\text{th}$ percentiles of the 100 shape parameter estimates after adding photon noise. 

In these simulations, pixel uncertainties are assumed to be independent and are drawn from the pipeline variance spectrum. 
This neglects covariance introduced by processing steps such as wavelength interpolation, convolution with the LSF, and continuum normalization, and may therefore slightly underestimate the true parameter uncertainties.

It is possible that fitting Voigt profiles to lines to account for asymmetries and damping in the wings may be a more physically realistic model as compared to the Gaussian profile \citep{gray_2005}. 
We also fit all lines with Voigt profiles with the inclusion of a damping parameter for Lorentzian broadening. 
However, the Voigt fits generally produced slightly larger reduced $\chi^2$ values than Gaussian fits, indicating that the additional model complexity (after adding an additional free parameter) was not justified. 
We will explore more complex models, including Voigt profiles, Gauss–Hermite series expansions \citep{holzer_2020, salzer_2025}, and bisector measurements, in future work. 
For the purposes of estimating the line center to obtain an accurate RV, a Gaussian model is a sufficient approximation.

\section{Vetting Lines that deviate from Weighted Time Average \label{Vetting Lines that deviate from Weighted Time Average}}

\begin{figure*}[htb]
    \centering
	\includegraphics[width=18 cm]{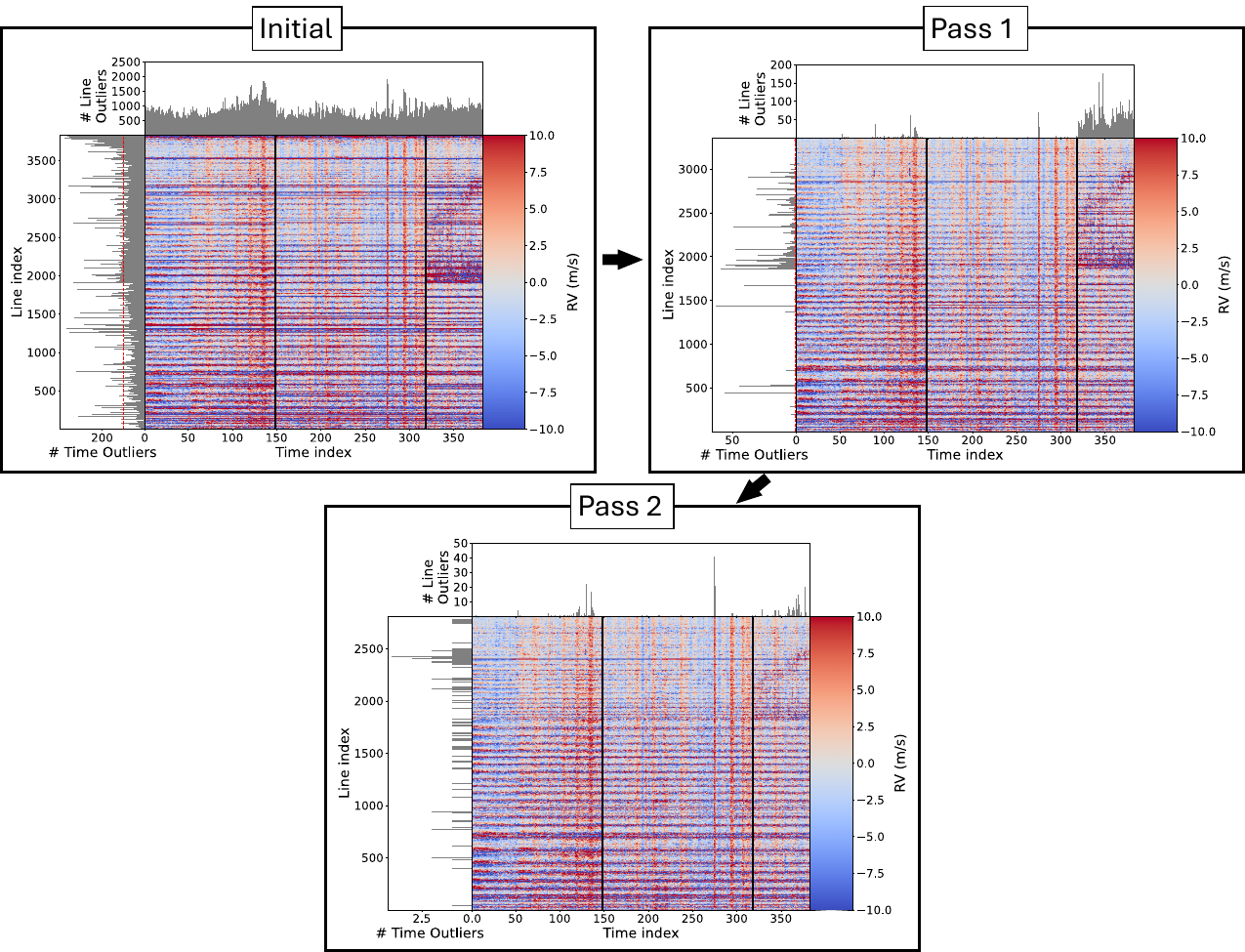}
    \caption{The procedure to remove lines that significantly deviate from the behavior of all lines and cannot be explained by photon noise, based on Section \ref{Line Vetting}. 
    \textit{Top left:} the initial heatmap of RVs for every line across time. 
    Vertical solid black lines in each heatmap indicate boundaries between observing runs (see Table \ref{tab:observation_vetting}). 
    The left panel of this figure shows, for each line, the number of days on which that line is an outlier. 
    Similarly, the top panel of this figure shows, for each day, the number of lines on which that day is an outlier. 
    See Appendix \ref{Vetting Lines that deviate from Weighted Time Average} on how these outliers are determined. 
    Lines with more than 100 time outliers (dashed red line) are removed in the first filtering pass. 
    \textit{Top right:} heatmap of RVs for every line across time after removing lines from the first filtering pass. 
    This removes many problematic lines in the reddest orders as well as many lines that have known wavelength calibration issues due to the LFC upgrade at the end of run 2. 
    For the second pass, we remove all lines that have more than 0 time outliers (with $\sigma_\text{thresh} = 20$), as shown by the dashed red line on the left. 
    This removes the remaining problematic lines. 
    \textit{Bottom:} heatmap of RVs for the remaining lines across time after the second filtering pass. 
    }
    \label{fig:vetting_combined}
\end{figure*}

In this section, we describe the procedure to reject lines whose RV measurements significantly deviate from the RVs of all lines and whose discrepancies cannot be explained by photon noise.
The top left panels of Figure~\ref{fig:vetting_combined} show the RV time series for all lines, along with the number of times each line is flagged as an outlier for each day, and the number of times each day is flagged as an outlier for each line.
Most lines have RVs between $-10$ to $+10$ m/s. 
Rotational modulation can be seen for some time intervals (time indicies $\approx 100-150$, corresponding to early 2022). 
The horizontal striping pattern arises from variations in SNR across the blaze functions of adjacent physical \'echelle orders. 
Since the lines are ordered approximately by wavelength, groups of neighboring lines alternately sample the high- and low-throughput regions of the blaze profiles, which produces the alternating banded pattern.

There are certain groups of lines that exhibit large RV deviations from the rest of the lines. 
This is most notable during run 3 (time indicies $320 - 383$) for lines between physical \'echelle orders $114 - 134$ (line indicies $2000 - 3000$). 
This is likely due to the updated algorithms to derive wavelength calibrations after the LFC was upgraded with a new PCF at the end of run 2, which caused some lines to show spurious RVs. 
There are also some lines in the reddest orders (line indicies $3750 - 3830$) whose RVs deviate significantly from the RVs of all other lines throughout all days. 

\begin{comment}
    \begin{figure*}
    \centering
	\includegraphics[width=18 cm]{Figures/Line Vetting/rv_heatmap_v1.pdf}
    \caption{A heatmap of RVs for every line across time. The left panel shows, for each line, the number of days on which that line is an outlier. Similarly, the top panel shows, for each day, the number of lines on which that day is an outlier. See Section \ref{Removing Lines that deviate from Weighted Time Average} on how these outliers are determined. Lines with more than 100 time outliers (dashed red line) are removed in the first filtering pass.}
    \label{fig:rv_heatmap_v1}
\end{figure*}
\end{comment}

An RV measurement for line $l$ at time $t$ is labeled an outlier if it satisfies
\begin{equation}
    |\mathrm{RV}^{\mathrm{wt}}_t - \text{RV}_{l,t}| \;>\; k_\text{thresh} \sqrt{\,{\sigma^{\mathrm{wt}}_t}^2 + \sigma_{l,t}^2}.
    \label{eq:outlier_condition}
\end{equation}
In other words, the measurement is considered an outlier if the deviation exceeds $k_\text{thresh}$ times the combined quadrature-summed uncertainty of the weighted mean and the individual line. 
Finally, outliers can be summed across individual lines, and lines that have a large number of time outliers can be removed from further analysis. 

We note that the above process can be repeated to remove days that have many lines as outliers. 
This would be done by computing a time-weighted average across lines $\mathrm{RV}^{\mathrm{wt}}_l$ and the corresponding uncertainty $\sigma^{\mathrm{wt}}_l$, checking how many lines were outliers for every day (based on Equation \eqref{eq:outlier_condition} but using the time-weighted average across lines), and removing days that have a large number of line outliers. 
However, as we will show, most of the problematic days can be explained by removing problematic lines. 
We prioritize keeping all of the days and removing lines because our line list is far larger than the number of days.  

For the first filtering pass, we use $k_\text{thresh} = 5$ and filter all lines that have more than 100 outlier days. 
This is shown in the left-most panel of the top-left figure of Figure~\ref{fig:vetting_combined}. 
This cutoff removes all of the lines in the reddest orders as well as many lines that have known wavelength calibration issues due to DRP v1.4 not having updated algorithms after the LFC was upgraded with a new PCF at the end of run 2. 

The top-right heatmap in Figure~\ref{fig:vetting_combined} shows the RV time series of only lines that are kept after the first pass. 
There are still some problematic lines in the red orders that have RVs that deviate significantly from the RVs from other lines during run 3, so we apply a second filtering pass with $k_\text{thresh} = 20$ and filter all lines that have more than 0 outlier days. 

\begin{comment}
\begin{figure*}
    \centering
	\includegraphics[width=18 cm]{Figures/Line Vetting/rv_heatmap_v2.pdf}
    \caption{Same as \ref{fig:rv_heatmap_v1}, but after removing lines that have more than 100 time outliers. For the second pass, we remove all lines that have more than 0 time outliers (with $\sigma_\text{thresh} = 20$), as shown by the dashed red line on the left.}
    \label{fig:rv_heatmap_v2}
\end{figure*}
\end{comment}

The bottom-left heatmap of Figure~\ref{fig:vetting_combined} shows the RV time series of only lines that are kept after the second pass. 
There are only a few problematic lines remaining and only a handful of days that still have a significant number of line outliers. 

\begin{comment}
\begin{figure*}
    \centering
	\includegraphics[width=18 cm]{Figures/Line Vetting/rv_heatmap_v3.pdf}
    \caption{Same as \ref{fig:rv_heatmap_v1}, but after removing the lines that have more than 0 time outliers with $\sigma_\text{thresh} = 20$.}
    \label{fig:rv_heatmap_v3}
\end{figure*}
\end{comment}

\section{LBL vs CCF RVs \label{LBL vs CCF RVs}}
In this section, we compare the RVs from the LBL approach in this work to the CCF approach based on \citet{ford_2024}. 
The weighted-mean LBL RV time series and corresponding uncertainties are determined using Equations \eqref{EQ:weighted_rv_timeseries} and \eqref{EQ:weighted_rv_uncertainty} using the curated line list determined in Section \ref{Line Vetting}. 
The run-dependent zero-point offsets were removed by subtracting the mean RV within each run.

We determine CCF RVs based on \citet{ford_2024} using the EchelleCCFs.jl package\footnote{\url{https://github.com/RvSpectML/EchelleCCFs.jl}}. 
First, we generate a top-hat mask based on the curated line list determined in Section \ref{Line Vetting} using a half width of $\Delta v = 1$ km~s$^{-1}$. 
We then evaluate the CCF on a velocity grid containing 121 velocities spanning $-$15 km~s$^{-1}$ to $+$15 km~s$^{-1}$ with a step size of 250 m~s$^{-1}$. 
The mask is used on the order-by-order daily-averaged spectra to get order-by-order RVs for each day. 
The run-dependent zero-point offsets are removed by subtracting the mean RV within each run. 
These order-by-order RVs are then combined using a weighted average, with weights proportional to the inverse variance of the order-by-order RVs. 

\begin{figure*}[hbt!]
    \centering
	\includegraphics[width=18 cm]{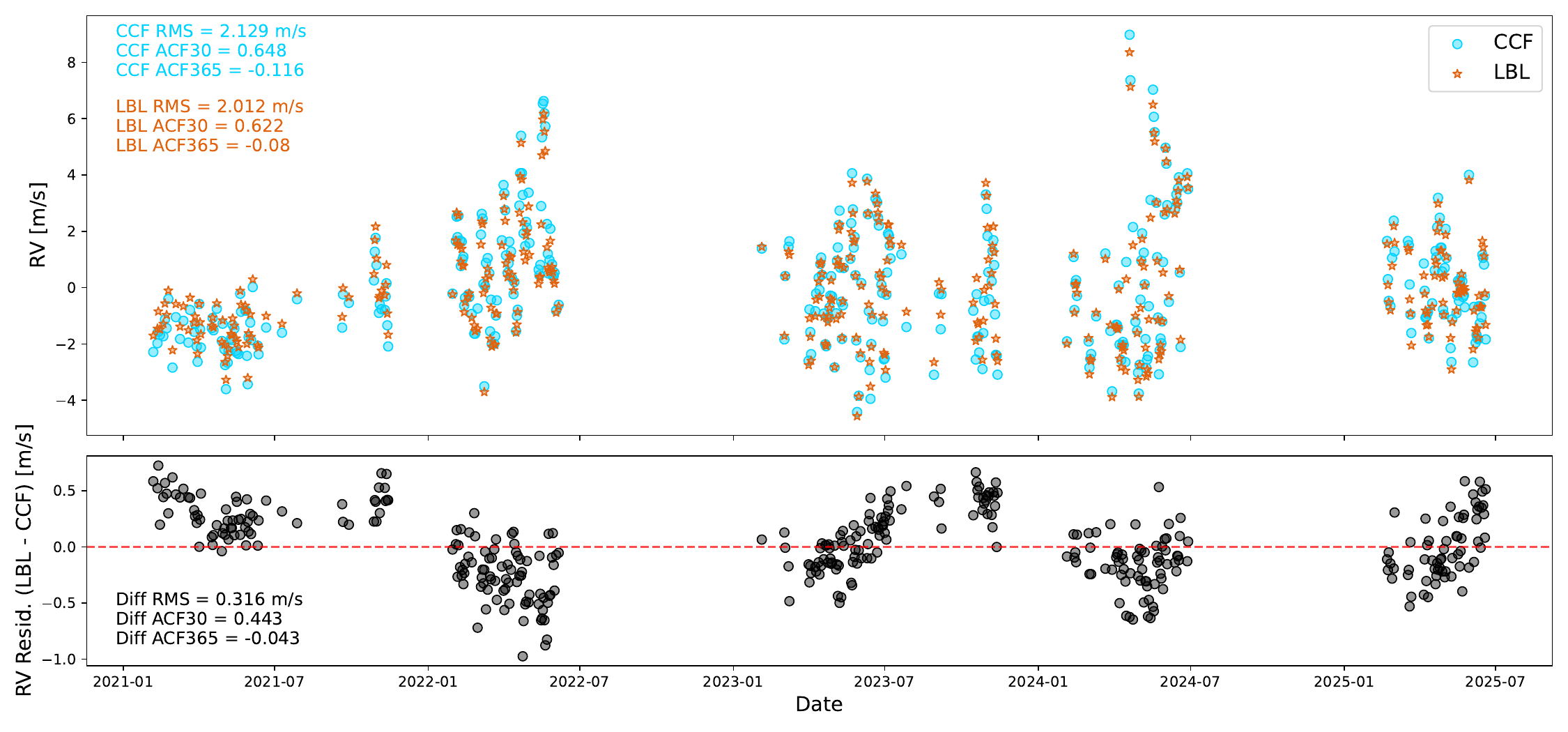}
    \caption{Comparison of RVs derived from the CCF method based on \citet{ford_2024} and the LBL method in this work. 
    \textit{Top}: the RV time series for the CCF (teal circles) and LBL (orange stars) methods. 
    The RV RMS of the LBL approach is slightly less ($\approx 2.012~\mathrm{m\,s^{-1}}$) than the CCF approach ($\approx 2.129~\mathrm{m\,s^{-1}}$), and also shows a lower periodicity at the solar rotation period and at 365 days. 
    \textit{Bottom}: the RV residuals between the LBL and CCF approach. 
    The difference in RVs measured by the two methods are mostly within $\approx 0.5~\mathrm{m\,s^{-1}}$. 
    }
    \label{fig:LBL_vs_CCF_timeseries}
\end{figure*}

\begin{figure}[hbt!]
    \centering
	\includegraphics[width= 8 cm]{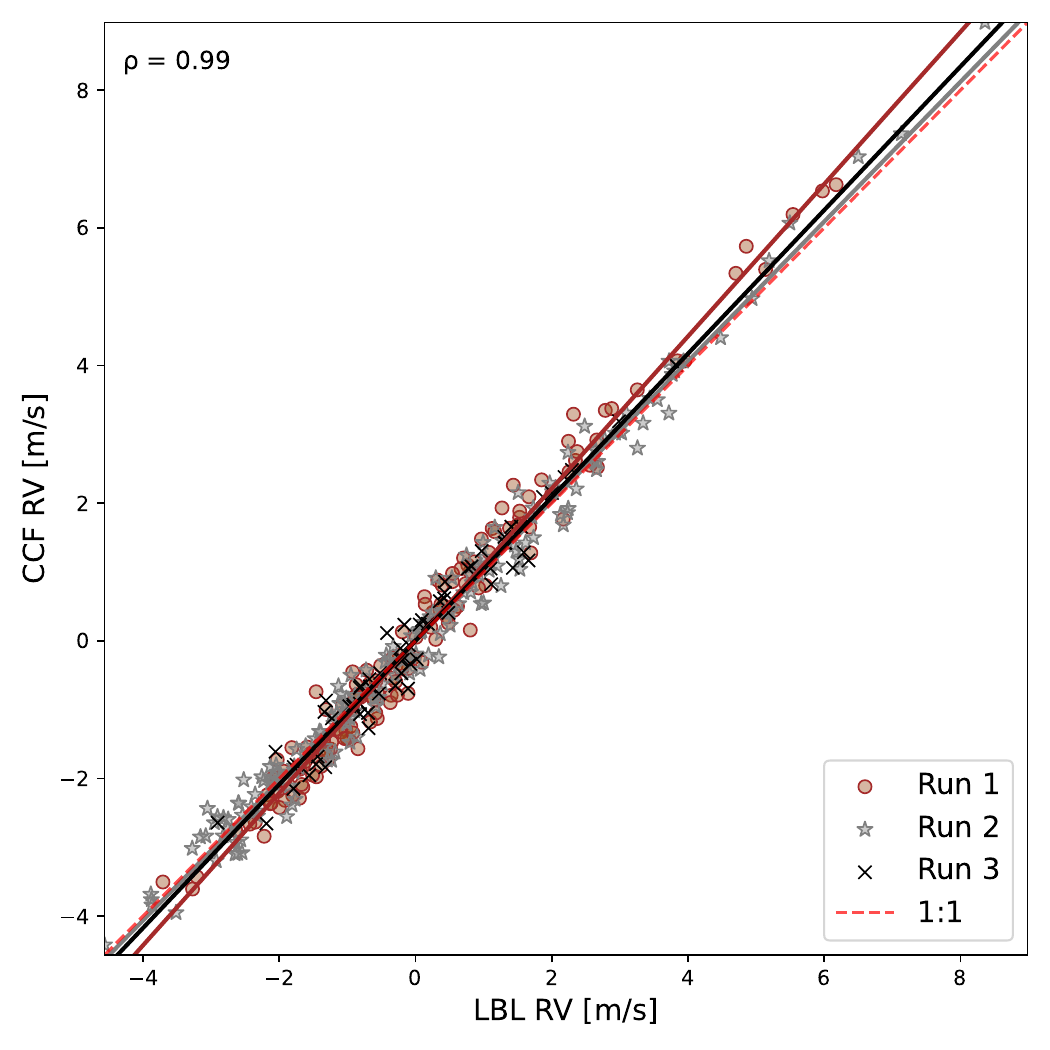}
    \caption{The CCF vs LBL RVs from runs 1 (brown circles), run 2 (gray stars), and run 3 (black X markers).  
    The CCF and LBL RVs show a strong correlation of $\rho = 0.99$. 
    Solid colored lines depict a best-fit linear slope to CCF vs LBL RVs within each run, and the red dashed line shows unity. 
    }
    \label{fig:rv_correlation_with_slopes}
\end{figure}

Figure \ref{fig:LBL_vs_CCF_timeseries} shows the RV time series of both the LBL and CCF methods, and Figure \ref{fig:rv_correlation_with_slopes} shows their correlation. 
The RV RMS of the LBL approach is slightly less ($\approx 2.012~\mathrm{m\,s^{-1}}$) than the CCF approach ($\approx 2.129~\mathrm{m\,s^{-1}}$). 
Since both methods use the same line list, the slight improvement in the LBL RVs could be due to the different weighting schemes used for each method. 
In particular, the LBL approach permits more fine-tuned weighting at the level of individual spectral lines, whereas the CCF approach permits weights at the order-by-order level. 

The LBL RVs show rotationally-modulated variations at the solar rotation period near the end of runs 1 and 2 for both the LBL and CCF methods. The autocorrelation function at a lag of the solar rotation period (ACF30; see Equation \eqref{EQ:ACF} and Section \ref{Autocorrelation and Activity Diagnostics of LBL RVs}) is $\approx$ 0.622, which is similar to ACF30 $\approx$ 0.648 for the CCF RVs.
In addition, the LBL RVs show that the autocorrelation function at a lag of 365 days (ACF365; see Equation \eqref{EQ:ACF} and Section \ref{Autocorrelation and Activity Diagnostics of LBL RVs}) is $\approx-0.08$, which is similar to ACF365 $\approx-0.116$ for the CCF RVs. 
The reduced power of ACF30 and ACF365 for the LBL method as compared to the CCF method could be due to the down-weighting of lines particularly susceptible to stellar activity, tellurics, or instrumental effects.

Figure \ref{fig:LBL_vs_CCF_timeseries} also shows that the differences between the LBL and CCF methods are largely within $\approx 0.5~\mathrm{m\,s^{-1}}$. The RMS of the difference between LBL and CCF RVs is $0.316~\mathrm{m\,s^{-1}}$, which is slightly lower than the value of $0.4~\mathrm{m\,s^{-1}}$ for the difference between HARPS-N data reduction software and the LBL RVs computed in \citet{dumusque_2018}. 

Figure \ref{fig:rv_correlation_with_slopes} shows that the LBL and CCF RVs are highly correlated with a Spearman Correlation Coefficient of $\rho = 0.99$, and the correlation is largely consistent within each run. 
To compare the consistency of the CCF and LBL measurements within each run, we fit linear slopes $b$ to the RVs within each run and find $b = 1.106, 1.015, 1.043$ for runs 1, 2, and 3, respectively. 
The slopes, particularly run 1, are slightly larger than unity, suggesting that CCF RVs are more susceptible to contamination from solar activity.

\section{Derivation of the Expected Jaccard Similarity \label{Derivation of the Expected Jaccard Similarity}}
In Section \ref{Monte Carlo Simulations of FLARES}, we computed the expected Jaccard similarity for two independently drawn random subsets of size $N$ from a total of $L=2809$ lines. 
Let $A$ and $B$ be two such subsets. 
For any given line, the probability that it is included in a subset is $p = N/L$. 
The expected size of the intersection is therefore
\begin{equation}
\mathbb{E}[|A \cap B|] = L p^2 = \frac{N^2}{L},
\end{equation}
since each line must be selected independently in both subsets.
Similarly, the expected size of the union is
\begin{equation}
\mathbb{E}[|A \cup B|] = \mathbb{E}[|A|] + \mathbb{E}[|B|] - \mathbb{E}[|A \cap B|]
= 2N - \frac{N^2}{L}.
\end{equation}
We therefore approximate the expected Jaccard similarity as
\begin{equation}
J_{\mathrm{rand}}(N) \approx \frac{\mathbb{E}[|A \cap B|]}{\mathbb{E}[|A \cup B|]}
= \frac{N^2/L}{2N - N^2/L}
= \frac{N}{2L - N}.
\end{equation}

\bibliography{example}{}
\bibliographystyle{aasjournal}

\end{document}